\documentclass[prb,superscriptaddress,showpacs,twocolumn]{revtex4}
\usepackage{graphicx,amssymb}
\begin{document}

\title{ Structural and magnetic properties of the single-layer manganese oxide 
  La$_{1-x}$Sr$_{1+x}$MnO$_4$ }

\author{ S. Larochelle }
\altaffiliation
{Present Address: Department of Physics, University
of Toronto, Toronto, Ontario M5S 1A7, Canada}
\affiliation{Department of Physics, Stanford University, Stanford, CA 94305}
\author{ A. Mehta }
\affiliation{Stanford Synchrotron Radiation Laboratory,
  Stanford Linear Accelerator Center, Stanford, CA 94309}
\author{ L. Lu}
\author{ P. K. Mang }
\affiliation{Department of Applied Physics, Stanford University, Stanford,
  CA 94305}
\author{ O. P. Vajk }
\altaffiliation
{Present Address: NIST Center for Neutron Research,
National Institute of Standards and Technology, Gaithersburg,
Maryland 20899}
\affiliation{Department of Physics, Stanford University, Stanford, CA 94305}
\author{ N. Kaneko }
\altaffiliation
{Present Address: National Institute of Advanced Industrial Science and Technology,
Tsukuba Central 2-2,
Tsukuba, Ibaraki 305-8568, Japan
}
\affiliation{Stanford Synchrotron Radiation Laboratory,
  Stanford Linear Accelerator Center, Stanford, CA 94309}
\author{ J. W. Lynn }
\affiliation{NIST Center for Neutron Research, National Institute of Standards and Technology, 
  Gaithersburg, MD 20899}
\author{ L. Zhou }
\altaffiliation
{Present Address:
Department of Biomedical Engineering,
Emory University,
Atlanta, GA 30322
}
\affiliation{Stanford Synchrotron Radiation Laboratory,
  Stanford Linear Accelerator Center, Stanford, CA 94309}
\author{ M. Greven }
\affiliation{Stanford Synchrotron Radiation Laboratory,
  Stanford Linear Accelerator Center, Stanford, CA 94309}
\affiliation{Department of Applied Physics, Stanford University, Stanford,
  CA 94305}
\date{\today}

\begin{abstract}

Using x-ray and neutron scattering, we have studied the structural and 
magnetic properties of the single-layer manganite La$_{1-x}$Sr$_{1+x}$MnO$_4$ 
($0 \le x < 0.7$). Single crystals were grown by the traveling-solvent floating-zone 
method at 18 La/Sr concentrations. The low-temperature phase diagram can be
understood by considering the strong coupling of the magnetic and orbital
degrees of freedom, and it can be divided into three distinct regions:
low ($x<0.12$), intermediate ($0.12 \le x < 0.45$), and high ($x \ge 0.45$)
doping. LaSrMnO$_4$ ($x=0$) is an antiferromagnetic Mott insulator,
and its spin-wave spectrum is well-described by linear spin-wave theory
for the spin-2 square-lattice 
Heisenberg Hamiltonian with Ising anisotropy. Upon doping, as the
$e_g$ electron concentration $(1-x)$ decreases, both the two-dimensional
antiferromagnetic spin correlations in the paramagnetic phase and the
low-temperature ordered moment decrease due to an increase of frustrating
interactions, and N\'eel order disappears above $x_c = 0.115(10)$. The magnetic
frustration is closely related to changes in the $e_g$ orbital occupancies
and the associated Jahn-Teller distortions. In the intermediate region,
there exists neither long-range magnetic nor superstructural 
order. Short-range-correlated structural ``nanopatches" begin to form 
above $x \sim 0.25$.
At high doping ($x \ge 0.45$), the ground state of La$_{1-x}$Sr$_{1+x}$MnO$_4$
exhibits long-range superstructural order and a complex (CE-type)
antiferromagnetic order which differs from that at low doping.  The
superstructural order is thought to arise from charge and orbital ordering
on the Mn sites, and for $x=0.50$ we conclude that it is of $B2mm$
symmetry.
For $x>0.50$, the superstructural order becomes incommensurate with the lattice,
with a modulation wavevector $\epsilon$ that depends linearly on the
$e_g$ electron concentration: $\epsilon = 2 (1 - x)$. 
On the other hand, the magnetic order remains commensurate, but loses its
long-range coherence upon doping beyond $x = 0.50$.

\end{abstract}


\maketitle

\section{Introduction}

The perovskite and perovskite-derived manganese oxides have attracted considerable interest over
the last decade as the richness of their phase diagrams has been uncovered.
\cite{Tokura00,Dagotto01,Salamon01} Phases of interest 
include a ferromagnetic metallic phase, a paramagnetic phase with short-range 
structural distortions as well as
several antiferromagnetic phases. The competition between the first two phases seems to 
play a key role in the colossal magnetoresistance (CMR) effect,
while the presence of the other 
phases has renewed interest in the spin-orbital coupling issue in transition metal oxides.

The ``single-layer" material La$_{1-x}$Sr$_{1+x}$MnO$_4$
is the $n=1$ end-member of the Ruddlesden-Popper family
La$_{n(1-x)}$Sr$_{nx+1}$Mn$_n$O$_{3n+1}$ of manganese oxides.
La$_{1+x}$Sr$_{1-x}$MnO$_4$ does not  
exhibit CMR \cite{Moritomo95, Bao96} and 
its layered structure results in strongly anisotropic transport properties.
\cite{Moritomo96} 
A magnetic/charge/orbital ordered phase is observed at low temperature at half doping ($x=0.50$).
\cite{Sternlieb96, Murakami98} 
The low dimensionality makes La$_{1+x}$Sr$_{1-x}$MnO$_4$ an interesting model
system for the study of the underlying doped MnO$_2$ planes
and for a comparison with results for the 
double-layer ($n=2$) and perovskite ($n=\infty$) manganites.
Furthermore, La$_{1-x}$Sr$_{1+x}$MnO$_4$ is a structural 
homologue of the cuprate La$_{2-x}$Sr$_{x}$CuO$_4$ and the nickelate La$_{2-x}$Sr$_x$NiO$_4$ 
for which, in part because of their low dimensionality, the issue of charge inhomogeneity has 
been central in recent investigations. \cite{Tranquada95,Chen93}
The most frequently studied composition of the single-layer manganites
is La$_{0.50}$Sr$_{1.50}$MnO$_4$ ($x=0.50$), although research also has
extended to other La/Sr doping ratios \cite{Moritomo95,Bao96,Wakabayashi01,Larochelle01} 
and to (Nd,Sr)$_2$MnO$_4$ and (Pr,Sr)$_2$MnO$_4$.
\cite{Moritomo97,Autret01,Kimura02,Nagai02}

At room 
temperature, La$_{0.50}$Sr$_{1.50}$MnO$_4$ 
has a highly symmetric body-centered tetragonal structure (space
group $I4/mmm$ \cite{Bouloux81a}) which becomes strongly distorted at low temperature. The 
low-temperature structural phase was first observed by electron microscopy, 
which revealed a 
$(\frac{1}{4},\frac{1}{4},0)$ wavevector modulation of the room temperature structure  
below $\sim$ 220 K. \cite{Moritomo95,Bao96} An anomaly in
the resistivity at this temperature indicates that
the structural transition is associated with charge ordering.
The existence of a 
distinct low-temperature
phase was confirmed in a neutron scattering experiment,\cite{Sternlieb96}
although it was concluded that this phase has a higher structural symmetry, 
with a  
$(\frac{1}{2},\frac{1}{2},0)$ modulation 
of the room-temperature structure. This 
experiment also established the presence of antiferromagnetic order below 110 K. 
Based on the model originally developed by Goodenough \cite{Goodenough55}
for the half-doped perovskite 
La$_{0.50}$Ca$_{0.50}$MnO$_3$ ($n=\infty$),
these results were interpreted as an indication of charge order in 
La$_{0.50}$Sr$_{1.50}$MnO$_4$.
Resonant x-ray scattering 
involves virtual excitations from core to valence states
and thus can probe anisotropies in the valence charge density.
A study of Mn $K$-edge resonant scattering from La$_{0.50}$Sr$_{1.50}$MnO$_4$ 
revealed a strong enhancement of the low-temperature superlattice 
peaks, which was interpreted as a confirmation of
orbital and charge order in this compound. \cite{Murakami98}
However, this interpretation has been controversial.
\cite{Ishihara98,Benfatto99,Elfimov99,Mahadevan01,Ishihara02} 
The $K$-edge resonance involves virtual excitations to the unoccupied
4$p$ levels, and thus is only an indirect probe of 3$d$ states, and 
the signal is expected to be sensitive to both the lattice distortion and 
the atomic configuration. 
Recent soft x-ray scattering experiments at the Mn $L$ edge 
\cite{Wilkins03,Dhesi04,Thomas03} might be better 
evidence of orbital ordering since they directly probe the 3$d$ states,
and it has been suggested that it is possible to spectroscopically 
differentiate 
cooperative Jahn-Teller distortions
of the Mn$^{3+}$ ions and direct (or spin-correlation-driven)
orbital ordering.\cite{Wilkins03,Dhesi04}
In another study, \cite{Huang04}
related linear dichroism results at the Mn $L$ edge were
interpreted as evidence that the $e_g$ orbital 
order in the CE phase invloves $d_{x^2-z^2}$ and $d_{y^2-z^2}$ orbitals
rather than the 
$d_{3x^2-r^2}$ and $d_{3y^2-r^2}$ orbitals, as is generally assumed.
However, the two ordering patterns share the same 
lattice symmetry. Finally, from a consideration of the number of allowed
Raman modes, the orthorhombic $Pbmm$ symmetry was proposed
for the low-temperature phase,\cite{Yamamoto00}
which would manifest itself in a 
$(\frac{1}{4},\frac{1}{4},0)$ modulation of the room-temperature structure.

The phase transition at $x=0.50$ also causes changes in the optical properties of the material.
Infrared absorption \cite{Calvani96} and reflectivity \cite{Jung00} measurements reveal that the 
optical gap increases significantly below the 
transition temperature. Moreover, the material becomes birefringent in the low-temperature 
phase \cite{Ishikawa99} as a result of the orthorhombic distortion caused by the orbital order.
The low-temperature phase is fragile and can be destroyed by magnetic fields of 
$\sim 25$ T. \cite{Tokunaga99b}  
Furthermore, 
photo-induced melting of the charge/orbital phase with exposure of the sample to 1.55 eV 
laser light has been reported.\cite{Ogasawara01}
We will present results in this paper that are 
indicative of a partial x-ray induced melting in the
ordered phase, 
similar to what was reported previously for Pr$_{0.30}$Ca$_{0.70}$MnO$_3$ \cite{Kiryukhin97}
and La$_{0.875}$Sr$_{0.125}$MnO$_3$. \cite{Kiryukhin99}

Compared to the efforts made to study the lattice distortions associated with the insulating 
behavior, relatively little is known about the magnetic properties of the single-layer 
manganites. Magnetization, \cite{Moritomo96} muon spin rotation  \cite{Baumann03} and neutron 
scattering \cite{Bouloux81b,Kawano88,Sternlieb96,Bieringer02} measurements have established 
that La$_{1-x}$Sr$_{1+x}$MnO$_{4}$ is an antiferromagnet near 
$x=0$, $x=0.5$ and $x=1$, with N{\'e}el temperatures of about 120 K, 110 K and 170 K,
respectively. At $x=0$ and $x=1$,
the magnetic structure is antiferromagnetic in the MnO$_2$ 
planes, with spins aligned parallel to the
stacking direction. At half-doping, on the other hand,
the magnetic structure is related to the complex CE 
magnetic phase of the perovskite manganites and is composed of two interpenetrating lattices
of (antiferromagnetically-coupled) ferromagnetic zigzag chains. 
The spins in this case lie within the MnO$_2$ planes.
Finally, we note that there have been reports \cite{Moritomo95,Park00,Hong01,Baumann03} 
of glassy magnetic behavior 
at low temperature between dopings of $x=0.20$ and $x=0.60$.

In a previous Letter, \cite{Larochelle01} we examined the nature of the $e_g$ 
electron order in
La$_{1-x}$Sr$_{1+x}$MnO$_4$. We briefly described the 
low-temperature phase at doping $x=0.50$ and reported on the observation of an incommensurate
lattice distortion for $x>0.50$. The present paper is organized as follows:
after describing our experimental methods in 
Sec. II, we present our magnetic neutron scattering results 
for the magnetic phase at low doping ($0 \le x \le 0.15$)
in Sec. \ref{sect_gmag}.
In Sec. \ref{sect_ce}, we then discuss the structural and magnetic properties of the 
low-temperature phase that is present in the doping range $0.45\le x< 0.7$.
Section \ref{sect_imr} describes the short-range structural and magnetic correlations observed in
the intermediate doping range $0.15<x<0.45$. Finally, 
Sec. \ref{sect_pd} presents the magnetic and structural phase diagram of 
(La,Sr)$_2$MnO$_4$ and a discussion of our results.

\section{Experimental methods}

In order to obtain single crystals, stoichiometric amounts of La$_2$O$_3$, SrCO$_3$ and 
MnO$_2$ (99.99\% purity or higher) 
were mixed and calcinated three times for thirty-six hours in an alumina
crucible at temperatures of 
1300-1360$^\circ$C, with intermediate grinding. The reacted and ground
powder was then pressed into
cylindrical rods and sintered for twelve hours in air at 1500-1580$^\circ$C. Crystals 
were grown from the ceramic rods using a four-mirror optical image furnace at a speed of 
6 mm per hour. Prior to the growth, the ceramic rods were rapidly sintered
inside the image furnace at 
about 85\% of the power needed to melt them.
A total of 24 crystals were grown at
18 different compositions, ranging from $x=0.00$ to $x=0.67$, which is close to the
solubility limit of $x\approx0.70$. \cite{Moritomo95,Bao96}
The growth atmosphere varied from argon
(partial pressure of oxygen of $\sim 10^{-6}$ bar) for $x=0.00$ compound up to 6 bar 
atmosphere of O$_2$ for samples with $x=0.67$. 

For the x-ray measurements, 
crystal pieces of about 4x2x1 mm$^3$ were cut from the boule and mounted 
inside a displex. Single crystal x-ray measurements were performed on beam line 7-2 at the 
Stanford Synchrotron Radiation Laboratory (SSRL). An energy of 14 keV was selected from the 
wiggler spectrum with a double-crystal Si(111) monochromator. 
Additional high-resolution powder x-ray 
scattering measurements were performed on SSRL beam line 2-1. Single crystal 
pieces were 
finely ground and the resulting powder was packed into a cavity in a silicon zero-background 
(510) mount. The samples were mounted and cooled in a transfer helium gas cryostat. The high
momentum resolution was 
achieved by using a Si(220) double-crystal monochromator and a Si(111) analyzer.
Single-crystal neutron scattering measurements were carrier out on the 
thermal triple-axis instruments of the NIST Center for Neutron 
Research (NCNR).

\begin{figure}[t]
  \centerline{\includegraphics[width=8.9cm]{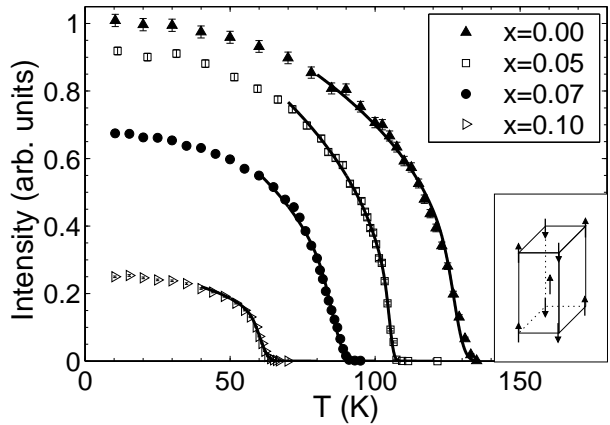}}
  \caption[Magnetic order parameter ($0\le x\le 0.10$)]
          {La$_{1-x}$Sr$_{1+x}$MnO$_4$ magnetic Bragg scattering intensity
           as a function of temperature,
            measured at the reciprocal space
            positions (1,0,4)$_m$ for $x=0.00$, (1,0,0)$_m$ for $x=0.05$,
            and (1,0,3)$_m$
            for $x=0.07$ and $0.10$. The subscript "$m$" indicates the
            unit cell of the magnetic structure shown in the inset.
            The intensity is normalized
            by the size of the crystals, and the low-temperature value for
            $x=0.00$ was set to 1.
            The lines are fits to the form
            $I \sim (1-T/T_{\rm N})^{2\beta}$ assuming a Gaussian distribution
            in N\'eel temperatures due to sample inhomogeneity.
            The values of $T_N$, $\beta$, and the ordered moment
	     are listed in Table \ref{momentstab}.
    \label{Neel}}
\end{figure}

\section{Magnetic Order and Spin Correlations  ($0\le \lowercase{x} < 0.15$)}
\label{sect_gmag}

The N\'eel phase of LaSrMnO$_4$ ($x=0$) was studied previously by several groups. 
\cite{Moritomo95,Baumann03,Kawano88,Bieringer02,Reutler03} These experiments found that 
LaSrMnO$_4$ orders antiferromagnetically with the K$_2$NiF$_4$ spin structure, an Ising 
anisotropy, and moments aligned along [0,0,1]. 
In four of the five experiments the N\'eel temperatures were determined to be
between 120 K and 130 K,
while $T_{\rm N}=180$ K was reported 
in Ref. \onlinecite{Kawano88}. 
In this Sec., we present neutron scattering measurements of the 
spin-wave dispersion for $x=0$, and of the order parameter
and instantaneous spin-spin correlation lengths for samples 
in the doping range 
$0\le x\le 0.15$. The goal of these measurements is to further establish
the properties of the magnetic phase of LaSrMnO$_4$ and to study its evolution upon doping.

Our results for the magnetic structure of LaSrMnO$_4$ are in good agreement with the earlier 
powder neutron diffraction studies. \cite{Kawano88,Bieringer02} 
The observed magnetic peaks can be indexed using the proposed structure, with 
nearest-neighbor (NN) moments antiferromagnetically aligned within the Mn-O plane, and with 
next-NN planes stacked 
ferromagnetically (see inset of Fig. \ref{Neel}). The structure is twinned, because the 
magnetic order of the intermediate Mn-O plane is not constrained by its two neighboring 
planes (the center Mn spin can point either up or down) and peaks corresponding to both 
twin domains were observed. No additional antiferromagnetically-stacked phase 
(observed, for example, in Rb$_2$MnF$_4$ \cite{Birgeneau70}) is discernible to a level of 
200 ppm. An antiferromagnetically-stacked phase would 
result in a Bragg peak with $L=0.5$ r.l.u., for example. 
No changes in the magnetic structure were noticed upon 
doping up to $x=0.10$.

\begin{table}
  \begin{center}
    \caption[N\'eel temperature, ordered moment, and order parameter critical exponent
      as a function of doping]
            {N\'eel temperature, ordered moment and order parameter critical exponent of
              La$_{1-x}$Sr$_{1+x}$MnO$_4$ ($0.00\le x\le 0.10$). The value
              of the ordered moment for $x=0.00$ is from Ref. \onlinecite{Bieringer02}.
              The values for other doping concentrations are derived from the relative values
              of the measured Bragg peak intensities,
              as discussed in the text.
            }
            \label{momentstab}
    \begin{tabular}{ccccc}
      \hline\hline
      $x$  &$T_{\rm N}$   &$M_{st}(x)/M_{st}(0)$&$M_{st}(x)$&$\beta$ \\
      \hline
      0.00 &128.4(5)&1              &3.3(2)$\mu_B$         &0.18(3)\\
      0.05 &105.5(2)&0.95(8)        &3.1(4)$\mu_B$         &0.20(3)\\
      0.075& 86.5(2)&0.82(6)        &2.7(4)$\mu_B$         &0.20(4)\\
      0.10 & 61.0(5)&0.50(4)        &1.6(2)$\mu_B$         &0.13(4)\\
      \hline\hline
    \end{tabular}
  \end{center}
\end{table}

The order parameter of samples in the antiferromagnetic phase was determined by measuring the 
temperature dependence of the intensity of one of the magnetic peaks, as shown in 
Fig. \ref{Neel}. 
The N\'eel temperature for the undoped sample is 128.4(5) K, in good agreement with four 
of the five previous measurements. \cite{Moritomo95,Baumann03,Bieringer02,Reutler03} 
The transition shows some 
rounding, most likely the result of chemical inhomogeneities from La-Sr substitutions. 
Allowing for a Gaussian distribution of the N\'eel temperature, 
we obtained
$\beta=0.18(3)$ for the order parameter critical exponent
of LaSrMnO$_4$ ($x=0$) from the fit shown in Fig. 1. 
This value of $\beta$
is similar to those of other two-dimensional Heisenberg 
antiferromagnets with an Ising anisotropy, and within two standard deviations of the 
2D Ising value $\beta=0.125$. \cite{Collins89} 
As discussed in detail below, the spin-wave spectrum features a relatively large Ising
anisotropy gap. Consequently, 
even though the rounding is not small, the extracted exponent is consistent with 
the presence of a significant Ising anisotropy. We note that the rounding $\Delta T_N$
(half-width-at-half-maximum) is $\Delta T_N/T_N = 1.5 - 4\%$ 
in the four samples with $x\le 0.10$. For example, for the $x=0.10$ sample, this
corresponds to doping inhomogeneities of less than $\Delta x = 0.003$.

As a next step, we studied the evolution of the antiferromagnetic order upon doping. 
In Fig. \ref{Neel}, 
the magnetic Bragg scattering intensities are scaled relative to each other based on 
data for the peak intensities that were
normalized per mole. The ordered staggered moment, $M_{st}(x)$, 
is proportional to the square root of the 
low-temperature intensity.
Only relative moments were measured in this experiment. Previous 
neutron powder scattering measurements indicate that the ordered moment at $x=0.00$ is 
$3.3(2) \mu_B$. \cite{Bieringer02} Establishing the absolute value of the ordered moments from 
single-crystal measurements requires a quantitative comparison
of the intensities of the magnetic Bragg peaks to those of nuclear peaks. Accurate measurements 
of nuclear peak intensities are difficult because the strong peaks generally suffer from extinction 
while the intensity of the weak nuclear peaks critically depends on the exact position of the atoms. 
One solution is to use small crystals which would limit the extinction of the strong nuclear 
peaks. This was not pursued as it would have hindered the correlation length measurements 
on the same samples 
(discussed below) for which the signal is much smaller. Another possibility would be to 
normalize the magnetic intensities by that of an acoustic phonon. 
As stated above, the ordered moment has been determined with good good accuracy for $x=0.00$.
\cite{Bieringer02} Given that the magnetic structure was found to stay the same up to 
$x=0.10$, we were able to pursue a third approach, the normalization of the magnetic 
intensities per unit volume for $0<x\le 0.10$ to the value determined previously for $x=0.00$.

\begin{figure}[t]
  \centerline{\includegraphics[width=8.5cm]{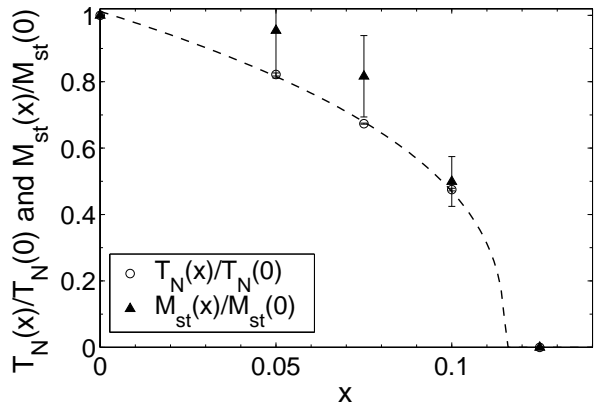}}
  \caption[N\'eel temperature and ordered moment as function of doping]
          {N\'eel temperature and ordered moment (measured at $T=10$ K) as function of doping
            for La$_{1-x}$Sr$_{1+x}$MnO$_4$. The values are normalized to those of the
            {$x=0.00$} sample. The dashed line is a guide to the eye for the doping dependence
	   of the N\'eel temperature, indicating our 
	   estimate of $x_c = 0.115(10)$ for the disappearance of two-dimensional long-range
	   order.
            \label{moments}}
\end{figure}

\begin{figure}[t]
  \centerline{\includegraphics[width=8.5cm]{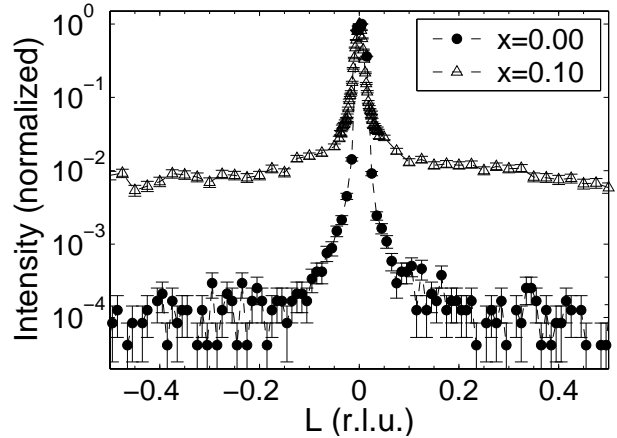}}
  \caption[L-scans]
	  {$L$-scans at $T=10$ K through the (1,0,0)$_m$ magnetic Bragg peak for $x=0$ and 0.10,
 	   normalized at the peak position.
           For $x=0$, the peak is resolution limited, indicating
	   three-dimensional long-range order, while it is broader than resolution for
	   $x=0.10$. 
Away from the
(1,0,0)$_m$ peak, the diffuse scattering is two orders of magnitude
higher for the $x=0.10$ sample. The extra intensity is due to a
rod of two-dimensional scattering. 
The data were taken on the spectrometer BT7 with 13.4 meV neutrons and
collimations of 35$^\prime$-40$^\prime$-sample-$25.8^\prime$-open.
	    \label{L-scans}}
\end{figure}

We observed (two-dimensional) long-range order up to $x=0.10$. 
Figure \ref{moments} shows the evolution of the N\'eel temperature and staggered ordered moment 
as a function of doping. We estimate that 
antiferromagnetic order disappears at 
$x_c=0.115(10)$. 
The ordered moment decreases approximately linearly with the 
N\'eel temperature. Table \ref{momentstab} reports the N\'eel temperature as well as the 
absolute and relative moments at $T = 10$ K. 
The magnetic Bragg peaks of LaSrMnO$_4$ are resolution limited in 
both in-plane and out-of-plane directions, indicating three-dimensional long-range order,
within the experimental precision. 
We note that the $x=0.10$ sample was not fully three-dimensionally
ordered at 10 K, the base temperature of our experiment, 
since Lorentzian broadening and a rod of two-dimensional scattering 
were discernible along $[0,0,1]$.
This is demonstrated in Fig. \ref{L-scans}.
Samples with higher doping ($x\ge 0.125$) only 
showed two-dimensional short-range correlations at 10 K.
The temperature dependence of the two-dimensional spin-spin correlation 
length in the paramagnetic phase will be described in detail below.

Reutler {\it et al.} \cite{Reutler03} 
reported long-range order for $x=0.125$. This discrepancy with our result
most likely stems from a difference in the oxygen content, since the crystal studied in 
Ref. \onlinecite{Reutler03} was grown under more reducing conditions (CO$_2$) than in the present 
study (air). From a comparison of the N\'eel temperatures, the oxygen 
stoichiometry difference between the two samples can be estimated to be about $\delta=0.015$.

\begin{figure}[t]
  \centerline{\includegraphics[width=9.0cm]{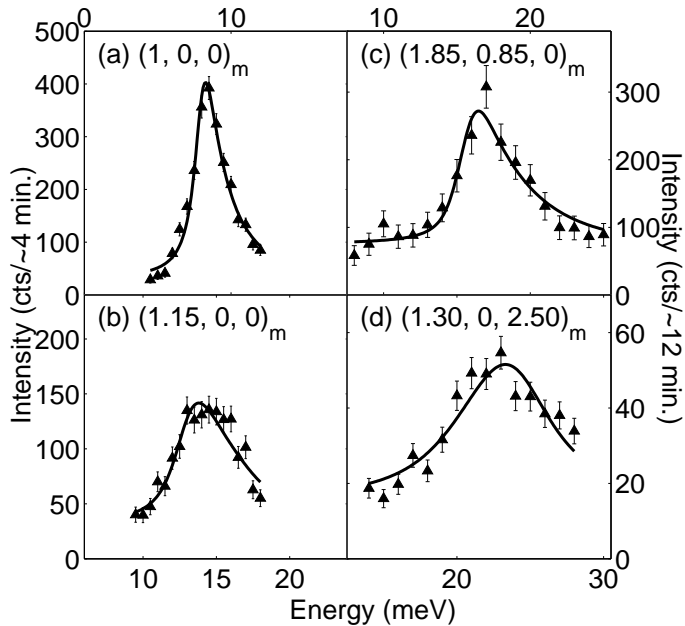}}
  \caption[LaSrMnO$_4$ spin waves at 10 K]
	  { LaSrMnO$_4$ spin waves at 10 K. The lines are fits of the spin-wave dispersion curve 
	    (assuming a Lorentzian cross section) convoluted with the resolution function 
	    of the neutron spectrometer.
	    The data were taken on the spectrometer BT2 with 
	    14.7 meV final energy neutrons and collimations of 
	    60$^\prime$-40$^\prime$-sample-40$^\prime$-80$^\prime$.}
	  \label{swscans}
\end{figure}

In order to establish the local magnetic parameters of LaSrMnO$_4$,
we measured the low-temperature spin waves
along the principal
crystallographic directions: $(\zeta,0,0)_m$, $(\zeta,\zeta,0)_m$, $(0.5,\zeta,0)_m$ and $(0,0,\zeta)_m$
(the reciprocal lattice considered in this Sec.
corresponds to the magnetic unit cell described
previously). We carried out fixed-momentum scans, varying the incident neutron energy.
Example scans are shown in Fig. \ref{swscans}. 
In order to fit the data, a Lorentzian
cross section was convoluted with the spectrometer resolution function using the program
ResLib. \cite{Zheludev01} Since the convolution depends on the shape of the dispersion curve,
the fit procedure was iterated several times, each time including the result for the dispersion
from the previous iteration. Convergence was reached after three iterations. The magnetic
excitations are not resolution limited and they broaden significantly at higher wavevectors,
from $\sim 1$ meV at the zone center to $\sim 4$ meV along the zone boundary.
The broadening above $\sim 20$ meV might be due to a coupling
between the spin-waves and optical phonons. A similar broadening was reported previously for the
double-layer compound (La,Sr)$_3$Mn$_2$O$_7$ and was attributed to phonon-magnon coupling.
\cite{Furukawa00,Hirota02} The excitations are dispersionless along $[0,0,1]_m$, as is
expected for a quasi-two-dimensional magnet.

\begin{figure}[t]
  \centerline{\includegraphics[width=9.0cm]{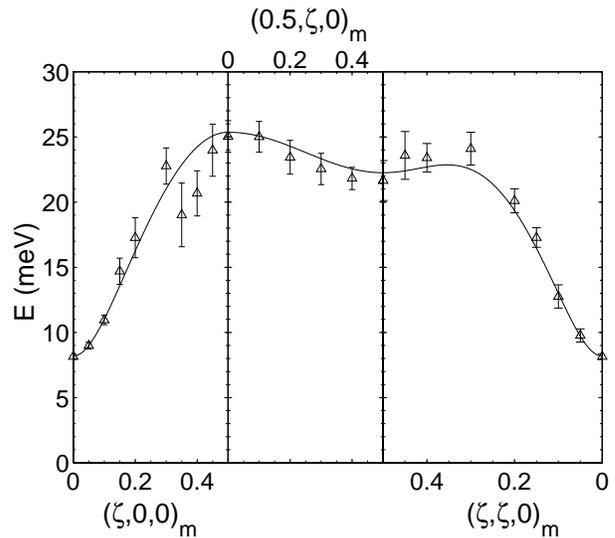}}
  \caption[LaSrMnO$_4$ spin-wave dispersion at 10 K]
      {LaSrMnO$_4$ spin-wave dispersion at 10 K. The line is Eq.
	(\ref{eq_disp}) with 
	parameters $J_1=3.4(3)$ meV, $\alpha=0.044(6)$ and $J_2/J_1=0.11(3)$.}
      \label{swdispersion}
\end{figure}

Our data for the spin-wave dispersion, summarized in Fig. \ref{swdispersion},
demonstrate that the magnetic degrees of freedom of LaSrMnO$_4$ 
are described to a good approximation by the two-dimensional square-lattice Hamiltonian

\begin{eqnarray}
  {\cal H}=J_1 \sum_{<i,j>_1} \left(S_i^x S_j^x + S_i^y S_j^y + 
          (1+\alpha) S_i^z S_j^z \right)  \nonumber\\
  + J_2 \sum_{<i,j>_2} \left (S_i^x S_j^x + S_i^y S_j^y + S_i^z S_j^z \right),\label{eq_hamilton}
\end{eqnarray}

\noindent where $J_1$ and $J_2$ are, respectively, the NN and 
second-NN
Heisenberg exchange couplings, and $\alpha$ is the Ising anisotropy. The sums run over first 
($\left< i,j \right>_1$) 
and second ($\left< i,j \right>_2$) nearest neighbors. 
The parameters $J_1$, $J_2$ and $\alpha$ of this Hamiltonian 
can be extracted by modeling the dispersion of the low-temperature spin-wave spectrum with the 
predictions from linear spin-wave theory.
For the Hamiltonian Eq. (\ref{eq_hamilton}), the spin-wave dispersion 
is given by

\begin{eqnarray}
  E({\bf q})&=&4SZ_cJ_1  \sqrt{\left(1+\alpha+\frac{1}{2}
    \frac{J_2}{J_1}\gamma_2({\bf q})\right)^2-\gamma_1({\bf q})^2} 
	\hspace{5mm} \label{eq_disp} \\
  \gamma_1({\bf q})&=&cos\left(\frac{1}{2}q_x a_m\right)cos\left(\frac{1}{2}q_y a_m\right) \nonumber\\
  \gamma_2({\bf q})&=&cos(q_x a_m)+cos(q_y a_m)-2,\nonumber
\end{eqnarray}

\noindent where $Z_c$ is a quantum renormalization factor (for spin 2, 
$Z_c\approx 1.04$). 
The NN coupling $J_1$ affects mostly the slope of the dispersion near the zone center, while 
the anisotropy $\alpha$ results in an energy gap at the zone center.
A non-zero
value of $J_2$ causes
a dispersion along the zone boundary (along $[\frac{1}{2},K,0]_m$).  
Overall, the dispersion is not very sensitive to higher-order corrections such as $J_2$, and
it is generally not possible to extract more than one independent higher-order parameter.
$J_2$ should thus be seen as an effective parameter that contains all the high-order 
(that is, beyond NN) contributions.
The three-parameter fit in Fig. \ref{swdispersion} provides a good description of our data,
and we obtain $J_1=3.4(3)$ meV, $\alpha=0.044(6)$ and 
$J_2/J_1=0.11(3)$. 
We note that for the $S=1/2$ NN square-lattice Heisenberg antiferromagnet
quantum fluctuations lead to a $\approx 7\%$ dispersion along the zone boundary.
\cite{Singh95,Kim01} In the $S=2$ case, however, quantum effects are expected to be 
significantly smaller, and the zone boundary dispersion observed for LaSrMnO$_4$ 
is dominated by the next-NN exchange $J_2$.

The value $\alpha = 0.044$ is quite large, and this results in a 
surprisingly large anisotropy gap of about 8 meV. 
The anisotropy is an order of magnitude larger than for model Heisenberg magnets 
such as Rb$_2$MnF$_4$ ($\alpha=0.0048$, Ref. \onlinecite{Cowley77}) or 
K$_2$NiF$_4$ ($\alpha=0.0021$,
Ref. \onlinecite{Birgeneau77}). However, it is still much lower than in systems that show nearly 
ideal two-dimensional Ising behavior, such as Rb$_2$CoF$_4$, where $\alpha=0.81(8)$. \cite{Ikeda78}

\begin{figure}[t]
  \centerline{\includegraphics[width=8.8cm]{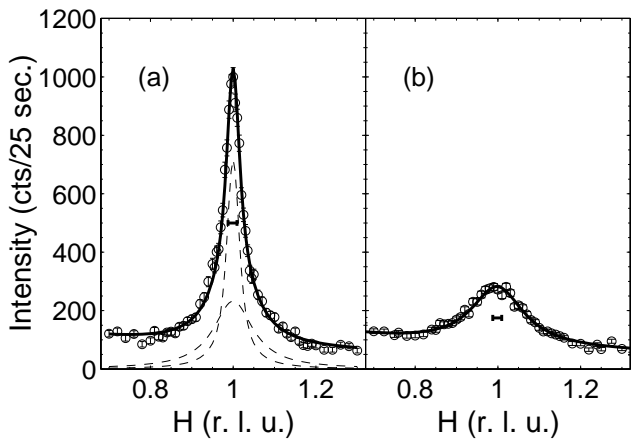}}
  \caption[Instantaneous magnetic structure factor measurement for LaSrMnO$_4$]
      {Examples of energy-integrating two-axis scans of the instantaneous magnetic structure factor 
	in the disordered phase of LaSrMnO$_4$. 
	Scan (a) was taken at $T=131.5$ K, in the temperature range just above the N\'eel 
	temperature ($T_{\rm N}=128.4(5)$ K) where the Ising anisotropy 
	is  important. The two dashed lines are the contributions from the 
	parallel (narrower peak) and perpendicular (broader peak) components. Scan (b) was taken 
	at $T=160$ K where the system behaves as an isotropic two-dimensional Heisenberg 
	antiferromagnet. The horizontal bars indicate the instrumental resolution.
	The data were taken on the spectrometer BT7 with 13.4 meV initial energy
	    neutrons and collimations of 35$^\prime$-40$^\prime$-sample-25.8$^\prime$-open.
	\label{scans}}
\end{figure}

Having gained a good understanding of the ordered phase, we next
measured the instantaneous antiferromagnetic correlations in the paramagnetic phase. 
In this phase, the magnetic scattering consists of rods at the antiferromagnetic wavevector 
positions in the $H-K$ plane. The scattering intensity is independent of the $L$ position
apart from a dependence on the magnetic form factor. Representative 
two-axis scans are shown in Fig. \ref{scans}. We extracted the correlation length by 
folding the instantaneous magnetic structure factor S({\bf q}) with the spectrometer 
resolution function. The structure factor has two components which correspond
to the spins fluctuations parallel and 
perpendicular to the easy-axis ([0,0,1]), as described in Ref. \onlinecite{Lee98}:

\begin{eqnarray}
  S({\bf q_{2D}})=sin^2(\phi)\frac{S_{||}(0)}{1+q^2_{2D}/\kappa^2_{||}} \nonumber\\
  + \left(1+cos^2(\phi)\right)\frac{S_\perp(0)}{1+q^2_{2D}/\kappa^2_\perp} \label{eq_SIsing}
\end{eqnarray}

\begin{figure}[t]
  \centerline{\includegraphics[width=9cm]{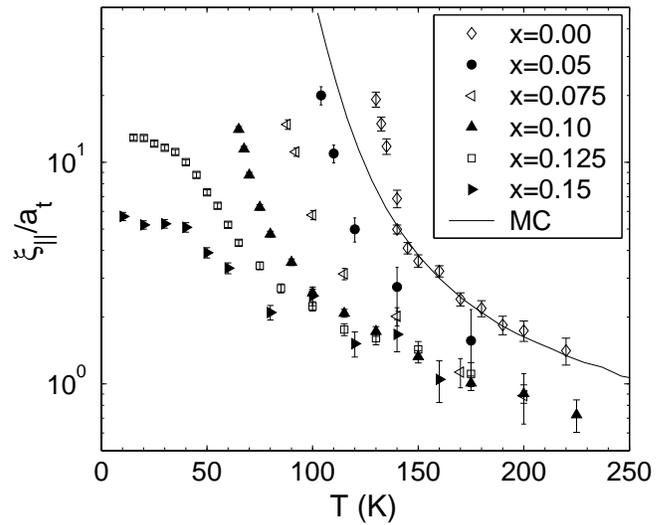}}
  \caption[Antiferromagnetic correlation length as function of temperature]
	  {Antiferromagnetic correlation length, in units of the tetragonal lattice constant $a_{t}$,
	    as function of temperature in 
	    La$_{1-x}$Sr$_{1+x}$MnO$_4$. The line is the result of 
	    a Monte-Carlo 
	    simulation of the spin-2 NN square-lattice Heisenberg 
	    antiferromagnet with an exchange coupling of $J=31.8$ K. 
	    \label{xi}}
\end{figure}

\noindent where $\kappa_{||(\perp)}$ is the inverse correlation length of the parallel 
(perpendicular) fluctuations, and $\phi$ is the angle between the scattering 
wavevector ${\bf Q}$ 
and the {\bf c} axis. The quantity $q_{2D}$ is the magnitude of the component of 
the reduced 
wavevector {\bf q} in the $H$-$K$ plane; ${\bf q}$ is defined within a Brillouin zone: 
${\bf q}={\bf Q}-{\bf G}$, where ${\bf G}$ is the nearest reciprocal lattice vector.
The $\phi$-dependent prefactors in Eq. (\ref{eq_SIsing})
originate from the fact that only the spin component perpendicular to the scattering wavevector
contributes to the magnetic cross section for unpolarized neutrons.
At temperatures well above the N\'eel temperature, the spin system is isotropic and 
the perpendicular and parallel 
components of the structure factor are indistinguishable. 
As the system is cooled toward the N\'eel temperature, the Ising anisotropy becomes relevant,
and the system undergoes a crossover from two-dimensional Heisenberg to two-dimensional Ising
behavior.
At the N\'eel temperature, the correlation length of the parallel component diverges while 
that of the perpendicular component remains finite. An estimate of this finite perpendicular 
correlation length is available within the framework of linear spin-wave theory:
\cite{Birgeneau77}

\begin{equation}
  \xi_{\perp}/a_m=\sqrt{Z_{c}/8\alpha}
\end{equation}

\noindent where $a_m$ is the in-plane lattice parameter of the magnetic unit cell.
Using the value of $\alpha$ obtained through the measurement of the spin-wave 
spectrum for the $x=0.00$ sample, $\xi_{\perp}=9.7(9)$ {\AA} at the N\'eel temperature. 
This value was used to fit the experimental data close to the N\'eel temperature since 
the fit was unstable if the parallel and perpendicular correlation lengths were both 
allowed to vary. The same value was used to analyze the data for $x=0.05$, 0.075 and 0.10.

Figure \ref{xi} shows the correlation length of the component of the magnetic 
fluctuations parallel to the easy-axis
for several dopings between $x=0$ and $x=0.15$. We also calculated the correlation
length for the spin-2 NN square-lattice Heisenberg antiferromagnet 
using the loop-cluster Monte-Carlo 
method. \cite{Evertz03}
We obtain good agreement (solid line) with the data at $x=0$ between 140 K 
and 250 K using an exchange coupling of $J=31.8$ K (2.75 meV). At temperatures lower than 140 K, 
the Ising anisotropy becomes relevant, and the measured 
correlation length diverges more strongly 
than for the Heisenberg model. 
The exchange coupling 
obtained from the comparison with the numerical result is somewhat smaller than the value
$J_1=3.4(3)$ meV obtained from the spin-wave dispersion curve. This discrepancy is probably 
due to the relatively strong frustrating second-NN coupling ($J_2\approx 0.11J_1$) 
which will lower the spin 
correlations and was not 
considered in the simulation.

\begin{table}[t]
  \begin{center}
    \caption[Effective spin stiffness and low-temperature correlation length for $x\le 0.15$]
	    {Effective spin stiffness and low-temperature correlation length for $x\le 0.15$. 
	      \label{xitab}}
    \begin{tabular}{lll}
      $x$ & \hspace{2mm} $\rho_{eff}$ (K) & \hspace{2mm} $\xi(x,0)/a_t$\\
      \hline
      0.00  & \hspace{2mm} 114(3) & \hspace{6mm} -- \\
      0.05  & \hspace{2mm} 91(5)  & \hspace{6mm} -- \\
      0.075 & \hspace{2mm} 72(5)  & \hspace{6mm} -- \\
      0.10  & \hspace{2mm} 51(4)  & \hspace{6mm} -- \\
      0.125 &  \hspace{4mm} --  & \hspace{2mm} 13.8(3) \\
      0.15  &  \hspace{4mm} --  & \hspace{2mm}  5.6(2) \\
    \end{tabular}
  \end{center}
\end{table}

For the spin-5/2 materials Rb$_2$MnF$_4$ and KFeF$_4$, it was found that the mean-field form
\cite{Keimer92}
\begin{equation}
\xi(T) = \frac{\xi_H (T)}{\sqrt{1 - \alpha \xi_H^2 (T)}},
\end{equation}
where $\xi_H (T)$ is the correlation 
length of the Heisenberg model, gives a good description of the crossover
from Heisenberg to Ising spin correlations. \cite{Lee98}
In the case of LaSrMnO$_4$, this form does not capture
the crossover, since the Ising anisotropy is an order of magnitude larger and, consequently, two-dimensional
Ising critical effects are relevant over a wider reduced temperature range. While the mean-field result
predicts a power-law exponent of $\nu=1/2$ for the divergence of the correlation length, the proper 
two-dimensional Ising result is $\nu = 1$. We note that our data are 
insufficient to extract the correlation length exponent from experiment.

Upon doping, the magnetic correlation length $\xi(x,T)$ decreases at any given temperature. 
For $x < x_c$, a possible heuristic description  
is to consider this decrease to be primarily due to a change of the 
spin stiffness. As Elstner {\it et al.} have shown,\cite{Elstner95}
the correlation length for the spin-$S$ NN square-lattice 
Heisenberg model falls on an approximately universal curve if one considers
$\xi(\rho/Z_{\rho})/(c/Z_c)$ as a function of $T/(\rho/Z_{\rho})$, where $\rho$ is the spin 
stiffness and $c$ is the spin-wave velocity.
Following this approach, and assuming that the quantum correction factors
$Z_\rho$ and $Z_c$ are unaltered from their $S=2$ values of $\sim 1$,\cite{Hamer94} 
we extract an effective spin stiffness. 
For example, at $x=0.10$ we estimate that $\rho_{eff}$ 
is about 40\% of the $x=0$ value. The values for $\rho_{eff}$ are reported 
in Table \ref{xitab}. 

The spin stiffness becomes zero at $x=x_c$.
For
the structurally related compound La$_{2-x}$Sr$_x$CuO$_4$, a simple empirical relation 
was found to hold for $x > x_c$:
\cite{Keimer92}

\begin{equation}
  \xi^{-1}(x,T)=\xi^{-1}(0,T)+\xi^{-1}(x,0)\, .
\end{equation} 

\noindent This form does not describe the present situation since
the correlation lengths for the short-range-ordered samples ($x=0.125$ and 0.15) exhibit
a significant temperature dependence already at intermediate temperatures.
For $x=0.125$ and 0.15, Table \ref{xitab} reports the 
estimated zero-temperature correlation length.

Magnetometry \cite{Moritomo95} and muon-spin-rotation \cite{Baumann03} measurements of 
La$_{1-x}$Sr$_{x+1}$MnO$_4$ suggest that the antiferromagnetic 
state near $x=0$ is replaced by a spin-glass phase at intermediate doping. The 
presence of a spin-glass phase points toward a frustration-induced suppression of the 
N\'eel order. The frustration appears as the system evolves toward the $x=0.50$ 
configuration in which the NN exchanges are ferromagnetic along zigzag 
chains and antiferromagnetic perpendicular to the chains.
It has been established from studies of the Ising model with random ferromagnetic and 
antiferromagnetic bonds ($\pm J$),\cite{Morgenstern80} and from studies of random mixtures
of magnetic ions with both ferromagnetic and antiferromagnetic couplings,
\cite{Matsubara84} that the presence of such random bonds in a 
(predominantly) NN magnet results in the destruction of the ordered magnetic 
ground state and its replacement by a spin-glass phase as the density of 
frustrating bonds increases.

\begin{figure}[t]
  \centerline{\includegraphics[width=9.0cm]{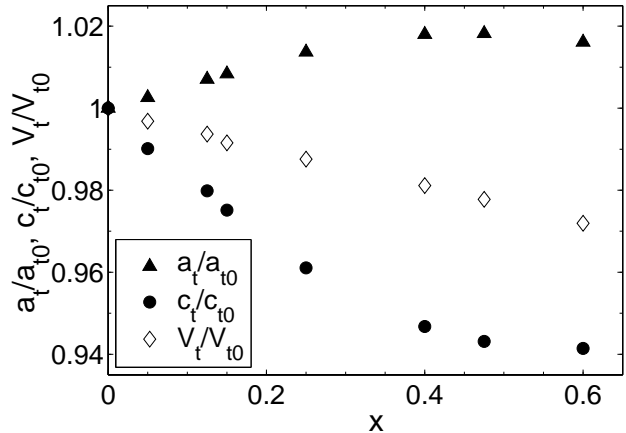}}
  \caption[Lattice parameters of La$_{1-x}$Sr$_{1+x}$MnO$_4$]
	  {Room-temperature lattice parameters and unit-cell volume of  La$_{1-x}$Sr$_{1+x}$MnO$_4$
	    for the tetragonal unit cell,
	    normalized to the $x=0$ values: $a_{t0}=3.796(4)$ \AA, 
	    $c_{t0}=13.18(2)$ {\AA}, and $V_{t0}=189.9(7)$ \AA$^3$. The unit-cell volume decreases linearly
	    with doping. The in-plane lattice parameter a$_t$ 
	    expands slightly while the out-of-plane 
	    lattice parameter $c_t$ contracts significantly with doping.
	    \label{latticep}}
\end{figure}

In the case of manganites, the sign of the exchange integral between NN
manganese ions can be related to the $e_g$ orbital occupations and to the associated 
Jahn-Teller distortion by considering the coupling between spins and orbitals (see, for 
example, Ref. \onlinecite{Tokura00}). We recall that the density 
of $e_g$ electrons, and hence of nominal Mn$^{3+}$ Jahn-Teller ions, is $1-x$.
In La$_{1-x}$Sr$_{1+x}$MnO$_4$,
at $x=0$, the Jahn-Teller distortions involve all Mn ions and are mostly along the $z$ direction 
($3z^2-r^2$ orbitals), while at $x=0.5$ most are within the $x-y$ plane ($3x^2-r^2$ and 
$3y^2-r^2$ orbitals \cite{Goodenough55} or $x^2-z^2$ and $y^2-z^2$ orbitals \cite{Huang04}) 
with half the sites (the ``Mn$^{4+}$'' sites) 
rather undistorted. The evolution of the lattice parameters indicates the trend of the
occupation probability of the out-of-plane
and in-plane Jahn-Teller orbitals. As shown in Fig. \ref{latticep},
our measurement of the room-temperature lattice parameters 
as a function of doping shows that the $c$ lattice constant decreases by 5.7\% between 
$x=0$ and $x=0.50$ while the $a$ lattice constant increases by 2\%. 
About 70\% of the change in the lattice parameters occurs between $x=0$ and $x=0.25$. 
In that range, the distribution of the Jahn-Teller orbitals should be rather random since 
no correlated distortion of the lattice is observable (see Sec. \ref{sect_imr}).

The location of the boundary between the antiferromagnetic and the spin-glass phases 
depends on the magnitude of the ferromagnetic ($J_F$) and antiferromagnetic ($J_{AF}$) couplings. 
From our spin-wave measurements for LaSrMnO$_4$, we estimate $J_{AF}$ is estimated to be 
3.4 meV ($J_1$).
However, the coupling
between Mn ions with different types of Jahn-Teller distortions could be different.
For the ferromagnetic coupling, an estimate can be obtained from the spin-wave 
measurements in the bilayer manganite (La,Sr)$_3$Mn$_2$O$_7$. \cite{Hirota02} Here, in the 
ferromagnetic state, the in-plane NN coupling is approximately $-5$ meV. Thus $J_F$ 
likely is comparable to $J_{AF}$ in the single-layer manganite. The situation for
La$_{1-x}$Sr$_{1+x}$MnO$_4$ can be compared to another compound with site-induced frustration: 
Rb$_2$Cu$_{1-x}$Co$_{x}$F$_4$. \cite{Dekker88} In that material, the coupling between 
copper sites is ferromagnetic ($J_{Cu-Cu}=-22.0$ K) and the other couplings are 
antiferromagnetic ($J_{Co-Co}= 90.8$ K  and $J_{Cu-Co} \approx 9$ K). The magnetic ground state 
of the compound is a ferromagnet in the range $0\le x\le 0.18$, an antiferromagnet for
$0.40\le x\le 1$ and a spin glass in the intermediate doping regime $0.18<x<0.4$. 
Since the antiferromagnetic phase 
disappears already at a lower critical doping level in (La,Sr)$_2$MnO$_4$ 
($x_c \approx 0.115(10)$), 
the frustration appears to be somewhat larger.

\section{The Charge-Ordered Phase ($0.45\le \lowercase{x}<0.70$)}
\label{sect_ce}

Many manganites at and near half doping exhibit a rather complex distortion of the 
lattice  accompanied by magnetic order. This low-temperature phase, 
usually referred to as the CE phase, appears to depend only 
weakly on the dimensionality of the system since it has been observed in the perovskite, the 
double-layer and the single-layer compounds. 
In this Sec., we discuss the distortion in the single-layer manganite 
La$_{1-x}$Sr$_{1+x}$MnO$_4$ in the doping region $0.45\le x<0.70$. 
We first present our results for 
the commensurate doping $x=0.50$ and
then extend the discussion 
to other doping levels.
Finally, we discuss the magnetic properties of the compounds in this region of the 
phase diagram.

\begin{figure}
  \centerline{\includegraphics[width=8.8cm]{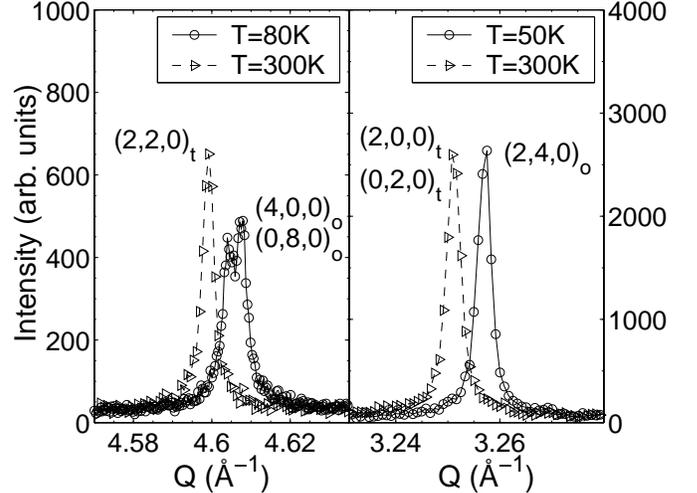}}
  \caption[Powder diffraction scans of La$_{0.50}$Sr$_{1.50}$MnO$_4$
    at high and low temperature]
	  {Comparison of La$_{0.50}$Sr$_{1.50}$MnO$_4$ powder diffraction scans at
	    room temperature and at low temperature. The scans show (a) the $(2,2,0)_t$ peak and 
	    (b) the $(2,0,0)_t$ peak. In the former case, the peak splits at low temperature,
	    indicating an orthorhombic distortion. In the latter case, the peak width 
	    stays constant, indicating the absence of a measurable monoclinic distortion.
	    The subscripts $t$ and $o$ denote the use of the tetragonal and orthorhombic 
	    unit cells, respectively. 
	    \label{monoortho}}
\end{figure}

\subsection{Structural Properties of LaSrMnO$_4$  ($x=0.50$)}

\begin{figure}
  \centerline{\includegraphics[width=8.5cm]{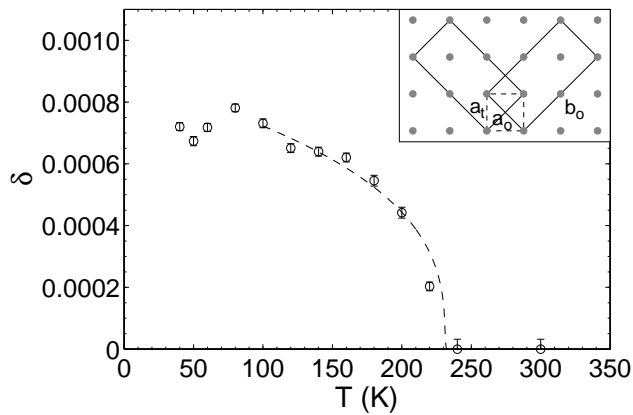}}
  \caption[Orthorhombicity as function of the temperature for La$_{0.50}$Sr$_{1.50}$MnO$_4$.]
	  {Orthorhombicity $\delta=|1-b_o/(2a_o)|$ as function of temperature 
	   for La$_{0.50}$Sr$_{1.50}$MnO$_4$. 
	   The structural transition occurs at $T_{COO} = 231.5(5)$ K, as determined from 
	   the temperature dependence of a superstructure peak (Fig. \ref{opc}).
	   The dashed curve is a guide to the eye. Inset:
	   Structural unit cells ($a$-$b$ plane)
           for $x=\frac{1}{2}$. The tetragonal high-temperature (orthorhombic low-temperature)
           unit cell is shown by dashed 
           (continuous) lines. At low temperature, there exist two twin domains that are rotated
           by 90$^{\circ}$ with respect to each other. Only the manganese sites are shown.
	    \label{orthor}}
\end{figure}

La$_{0.50}$Sr$_{1.50}$MnO$_4$ undergoes a structural phase transition at 
approximately 230 K. 
The low-tem\-pera\-ture phase is characterized by superlattice reflections 
with wavevector
$(\frac{1}{4},\frac{1}{4},0)_t$.
\cite{Moritomo95,Bao96,Larochelle01,Murakami98} 
The symmetry of the lattice can be established from 
extinction rules and high-resolution powder diffraction analysis. Powder diffractometry is 
a very reliable and sensitive probe of small lattice distortions because 
peaks corresponding to an identical $d$-spacing merge irrespective of the angular orientation,
and hence very small changes in the $d$ spacing are readily visible.
At room temperature, the powder diffraction peaks of
La$_{0.50}$Sr$_{1.50}$MnO$_4$
can be indexed on a tetragonal lattice with
lattice parameters $a_t \approx 3.86$ {\AA} and $c_t \approx 12.42$ {\AA}. 
However, as demonstrated in Fig. \ref{monoortho}, below the transition 
temperature, the $(2,2,0)_t$ peak splits into two, 
but the $(2,0,0)_t$ peak does not split or broaden. This indicates that the lattice becomes
orthorhombic with a 45$^\circ$ rotation of the axes in the $a$-$b$ plane (a broadening of the 
$(h,0,0)_t$ type reflections would indicate a further monoclinic distortion). The 
orthorhombicity varies as function of the temperature, as shown in Fig. \ref{orthor}.
The low-temperature and high-temperature unit cells are illustrated in the inset of Fig. \ref{orthor}. 

\begin{figure}
  \centerline{\includegraphics[width=9.0cm]{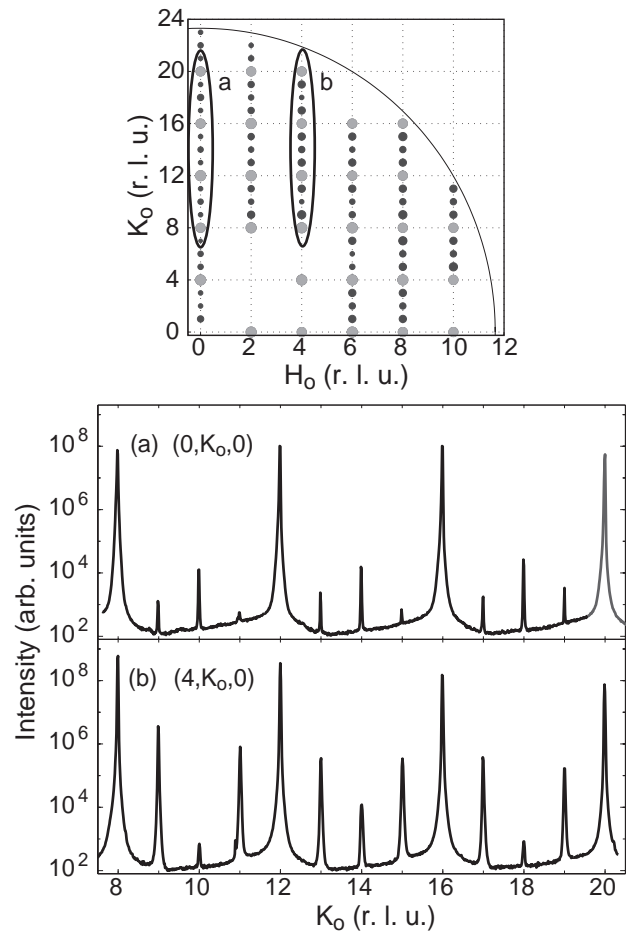}}
  \caption[Intensity map of the Bragg reflections at 100 K for La$_{0.50}$Sr$_{1.50}$MnO$_4$]
      {(Top) Intensity map of the Bragg reflections in the $L=0$ plane at 100 K for 
	La$_{0.50}$Sr$_{1.50}$MnO$_4$. The gray circles represent the high-symmetry Bragg peaks
	of the $I4/mmm$ structure, and the black circles represent the additional superlattice
	Bragg peaks of the low-temperature phase. The radius of the circles 
	is proportional to the logarithm of the intensities. The reciprocal lattice 
	corresponds to the low-temperature orthorhombic cell with $a_o=5.46$ {\AA},
	$b_o=10.92$ {\AA}, and $c_o=12.4$ {\AA}. The line (quarter circle) represents
	the maximum reachable momentum at the x-ray energy of 14 keV used in our experiment.
	 Several reflections,
	for example $(2,1,0)_o$, could not be reached due to geometric constraints.
	This map should be considered qualitative since absorption and extinction are considerable 
	in this material for 14 keV x-rays. 
	(Bottom) Two specific scans are shown, corresponding to the two regions
	labeled (a) and (b) on the map.
	\label{map}}
\end{figure}

\begin{figure}
  \centerline{\includegraphics[width=8.3cm]{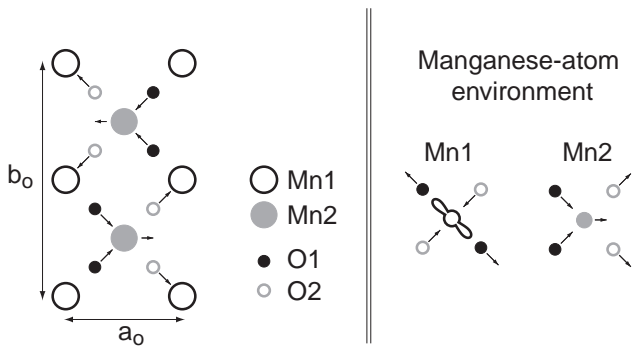}}
  \caption[Low-temperature distortion for $x=0.50$]
	  {Low-temperature distortion of the Mn-O plane with space-group symmetry $Bbmm$. 
	    There are two manganese sites and two basal oxygen sites in this space group.
	    The Mn1 (Mn2) sites are occupied by ``Mn$^{3+}$'' (``Mn$^{4+}$'') ions.
	    The panel on the right shows the distortion around each  
	    type of Mn site. The actual $B2mm$ symmetry of La$_{0.5}$Sr$_{1.5}$MnO$_4$ is lower
	    than that indicated here, with three rather than two inequivalent manganese sites,
	    as discussed in the text.
	  \label{Bbmm}}
\end{figure}

The symmetry of the low-temperature phase for $x=1/2$
was investigated further through a thorough survey of the reciprocal 
space of a single crystal. The reciprocal space map shown in Fig. \ref{map} exhibits the 
extinction symmetry $B${-}{-}{-}. A particularity of the low-temperature structure 
is the weakness of the reflections with $H_o=0$ and $K_o$ odd. These reflections are two to
three orders of magnitude smaller than equivalent reflections with $H_o$ different from zero.
This suggests that the crystal has a pseudo-symmetry $b$-glide and should thus be a subgroup
of the $Bbmm$ space group. $B2mm$ is the sole orthorhombic space group with the proper 
extinction symmetry,
and is thus the most probable space group.
A schematic of the distortion with the space-group symmetry $Bbmm$ is shown in Fig. \ref{Bbmm}.

Because of severe x-ray absorption and extinction effects, we
were not able to accurately determine the structure factors for the reflections, and hence 
were unable to calculate the atomic displacements to further establish the nature of 
the distortion.
Nevertheless, a few characteristics of the partially-determined structure are worth 
pointing out. A first characteristic is the presence of a mirror plane within the Mn-O layer,
perpendicular to [0,0,1].
Its presence forbids any buckling of the Mn-O plane.
Second, as a result of the low $B2mm$ symmetry, there exist three distinct manganese sites with a ratio of 
manganese atoms of 2:1:1. The single-multiplicity Mn sites are related by the 
pseudo-glide plane. Finally, the inclusion of the pseudo-glide plane in the set of symmetry
operators generates the supergroup $Bbmm$, in which the lattice distortion resulting in the
low-temperature structure is a shear-type distortion rather than a breathing-type 
distortion. A breathing-type distortion would have a mirror plane perpendicular to [1,0,0]$_o$
and would have $Bmmm$ symmetry.

The symmetry proposed here for the low-temperature phase of La$_{0.5}$Sr$_{1.5}$MnO$_4$
is in good agreement with previous electron diffraction results,
\cite{Moritomo95,Bao96} but it is lower than that reported in 
Ref. \onlinecite{Sternlieb96} based on neutron scattering measurements. We confirmed, through neutron 
scattering measurements, that the Bragg reflections with wavevector 
$(\frac{1}{4},\frac{1}{4},0)_t$ result from the bulk of the sample, and thus our neutron 
scattering data are in good agreement with our x-ray results. Finally, the 
symmetry identified by Raman scattering is $Pbmm$. \cite{Yamamoto00} This is a subgroup
of the pseudo-symmetry identified here ($Bbmm$) which differs only in the centering of the 
lattice (primitive versus face centered). The symmetry of the atomic sites in both space 
groups is the same. Consequently, both have the same Raman active modes. 
Therefore, $Bbmm$ is an acceptable symmetry to model the Raman modes. The breaking of the symmetry 
from $Bbmm$ to $B2mm$ is probably too weak to be observable as additional Raman modes.

Based on a local density of states calculation, it was concluded \cite{Mahadevan01} 
that the symmetry of the lattice structure is 
$Bbmm$, the 
pseudo-symmetry obtained experimentally. The calculation established that half
of the manganese sites have a Jahn-Teller-type elongation of the surrounding oxygen 
octahedron (the 
so-called ``Mn$^{3+}$'' sites) while the octahedra around the other sites are much less 
distorted (``Mn$^{4+}$'' sites). Furthermore, it was concluded that the difference in 
valence between the two sites should be relatively small. Unfortunately, the present 
scattering analysis can not solve this issue since the determination of the Mn valences
requires the knowledge of the Mn-O bond distances (bond valence sum). A major 
difference between the crystal structures derived from the LDA calculation 
(space group $Bbmm$) and from experiments ($B2mm$) is that, in the former, there 
are only two unique manganese sites, while in the latter, there are three. In both
structures, there is only one ``Mn$^{3+}$'' site. However, our diffraction measurements
imply the existence of two crystallographically distinct ``Mn$^{4+}$'' sites. From our 
measurements (from the 
strengths of the $(h,0,0)_o$ reflections), we conclude that the difference between the two 
``Mn$^{4+}$'' sites is small. 

The mostly shear-type distortion of the planar oxygen sublattice in the $Bbmm$ 
structural model makes the resonant terms of the x-ray scattering tensor around 
the Mn sites highly anisotropic, as shown in Ref.
\onlinecite{Mahadevan01}, for example. The magnitude and the symmetry of the calculated 
anisotropy is in complete agreement with 
previous resonant scattering measurements,
\cite{Murakami98} again confirming the validity of the proposed 
symmetry for the low-temperature phase of La$_{0.5}$Sr$_{1.5}$MnO$_{4}$.

A number of perovskites and double-layer manganites at or near $x=0.50$ have 
closely related low-temperature phases. Structural refinements based on neutron and x-ray 
powder scattering as well as single-crystal neutron scattering are available for several 
compounds: La$_{0.5}$Ca$_{0.5}$MnO$_{3}$, \cite{Radaelli97} Nd$_{0.5}$Sr$_{0.5}$MnO$_{3}$,
\cite{Woodward99}, LaSr$_2$Mn$_2$O$_7$
\cite{Argyriou00} and Pr$_{0.6}$Ca$_{0.4}$MnO$_{3}$.\cite{Daoud02}
Consistent with our observations for La$_{0.5}$Sr$_{1.5}$MnO$_{4}$,
all these materials were found to have a unit cell with $a$-$b$ plane dimensions 
$\sim \sqrt{2} a_t \times 2\sqrt{2} a_t$, with $a_t$ the NN
Mn-Mn distance. Three of the structures also share the basic feature that half the manganese 
sites exhibit a strong Jahn-Teller-type distortion while the oxygen octahedra surrounding 
the other manganese sites are nearly undistorted. The structure of the fourth 
compound, Pr$_{0.6}$Ca$_{0.4}$MnO$_{3}$, has been argued to be quite different, \cite{Daoud02}
with manganese sites that all have equivalent short and long Mn-O bonds, and a distortion  
due to a formation of molecular units of Mn-O-Mn. However, recent resonant
x-ray diffraction work is conistent with inequivalent Mn atoms that order in a
CE-type pattern.\cite{Grenier04}
Therefore, at least three of the above four $n=2$ and $n=\infty$ compunds
have distortions 
that are compatible 
with the $B2mm$ symmetry proposed here for 
the low-temperature structure of La$_{0.5}$Sr$_{1.5}$MnO$_{4}$.

Based on bond valence sums, the difference between the crystallographically distinct Mn ions 
in Pr$_{0.6}$Ca$_{0.4}$MnO$_{3}$ is insignificant.\cite{Daoud02} On the other hand, bond valence sums for 
the other three compounds show some degree of 
charge disproportionation between the two manganese 
sites: 3.5 and 3.9 for La$_{0.5}$Ca$_{0.5}$MnO$_{3}$; \cite{Radaelli97} 3.49 and 3.98 for
Nd$_{0.5}$Sr$_{0.5}$MnO$_{3}$; \cite{Woodward99} 3.67 and 3.87 for LaSr$_2$Mn$_2$O$_7$. \cite{Argyriou00}
La$_{0.5}$Sr$_{1.5}$MnO$_{4}$ has a similar low-temperature structural distortion
and, thus, it is probable that it also exhibits some degree of charge disproportionation,
although with a magnitude that can be expected to be significantly less than a whole electron. 
As discussed below, charge 
disproportionation on the Mn sites in the single-layer compounds might help explain the 
experimental results obtained for $x>0.50$.

\begin{figure}
  \centerline{\includegraphics[width=8.6cm]{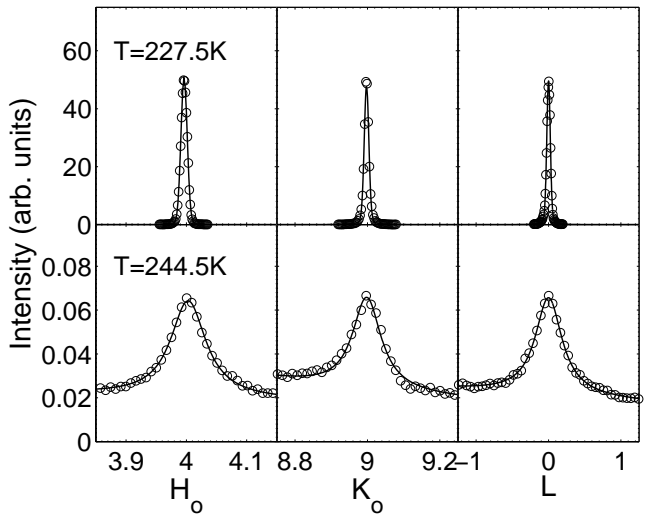}}
  \caption[Representative momentum scans through superlattice peak]
	  { Representative $H_o$, $K_o$ and $L$ scans of the $(9,0,0)_o$ superlattice peak 
	    of La$_{0.5}$Sr$_{1.5}$MnO$_4$.
	    The upper (lower) panels show scans below (above) the N\'eel temperature.\label{hkl}}
\end{figure}

\begin{figure}
  \centerline{\includegraphics[width=8.8cm]{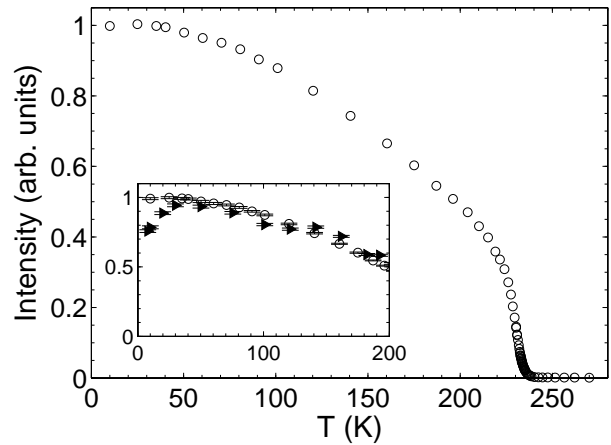}}
  \caption[Superlattice Bragg peak intensity as function of the temperature]
	  {Linear plot of the $(4,9,0)_o$ Bragg peak intensity in the ordered phase for $x=0.50$ as 
	    function of the temperature. The transition temperature is found to be 231.5(5) K. 
	Inset:
	Integrated intensity of the $(4,9,0)_o$ superlattice reflection for two different levels
            of x-ray flux on the sample. The high flux level
	    is $\sim 300$ times larger than the
            low flux. A partial melting of the ordered phase can be observed at low temperature
            for the data collected with the higher flux (black triangles).
	    \label{opc}}
\end{figure}

We also investigated the temperature dependence of the superlattice 
reflections. Below the transition temperature, the superlattice
peak $(9,0,0)_o$ was scanned along the three principal orthogonal axes and
the integrated intensity was obtained by modeling the profiles
with pseudo-Voight functions. Representative scans are shown in the top panels
of Fig. \ref{hkl} and the integrated intensity is displayed in 
Fig. \ref{opc}. Near the transition temperature, the data 
exhibit a rounding of about 3 K, which was too large to allow
a meaningful critical scattering analysis. Heat
capacity measurements on a piece from the same crystal boule also
showed broadening, in this case of about 4 K.
The broadening is most likely due to La/Sr inhomogeneities.
The transition is probably second order, since we found no signs of hysteresis,
and because there exist large fluctuations
with continuously varying correlation lengths above the charge/orbital
ordering temperature $T_{COO} = 231.5(5)$ K.

In the ordered phase of La$_{0.5}$Sr$_{1.5}$MnO$_4$, the peak widths along [1,0,0]$_o$
and [0,1,0]$_o$ increased continuously as the temperature was lowered
while it remained constant along [0,0,1]. This increase
is associated with an increase
of the crystal mosaic as the lattice becomes increasingly more
orthorhombic and, hence, more strained (the correlation between
the mosaic and the [1,0,0]$_o$ and [0,1,0]$_o$ directions resulted from the
scattering geometry of the experiment). The widths of the superlattice reflections
in all three directions were broader than those of the high-symmetry 
peaks, with correlations of about 200 {\AA} along $[0,0,1]_o$
and at least 1000 {\AA} along $[1,0,0]_o$ and $[0,1,0]_o$.

Above the transition, the superlattice peaks broadened considerably.
The bottom panels of Fig. \ref{hkl} show representative scans along the
three principal orthorhombic directions at $T=244.5$ K.
The scattering intensity was modeled as a convolution of the 
correlation function and the resolution function. An effective 
resolution function was defined from the peak shape of a nearby high-symmetry
peak, $(4,8,0)_o$, at a temperature of 232 K, just above $T_{COO}$.
A pseudo-Voight form was fitted 
in all three directions. Consequently, this resolution 
function included the crystal mosaicity in addition to the
fundamental instrumental resolution.
The correlation function used was a Lorentzian:

\begin{equation}
  C({\bf q})=\frac{C(0)\kappa^2}{\kappa^2-({\bf q}-{\bf G})^2 }
\end{equation}

\noindent where $C(0)$ is the amplitude, ${\bf G}$ is the position of the Bragg 
peak, and $\kappa$ is
the inverse correlation length.

\begin{figure}[t]
  \centerline{\includegraphics[width=8.8cm]{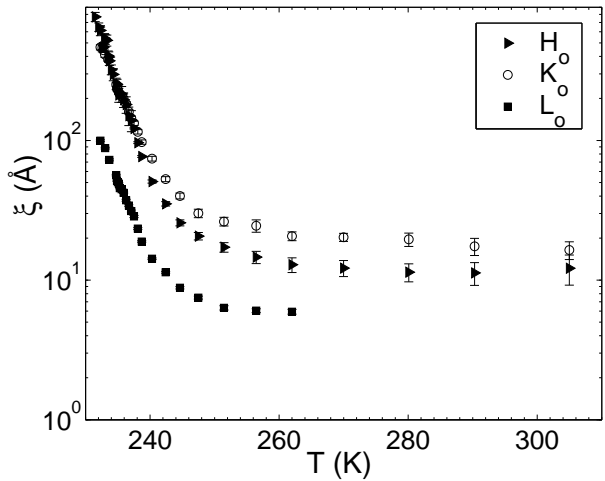}}
  \caption[Correlation length of the charge/orbital order as function of temperature]
	  {Logarithmic plot of the correlation length of the charge/orbital order as a function 
	    of temperature for $x=0.50$. Short-range distortions associated with the low-temperature phase 
	    are still present at room temperature. The correlation length is nearly constant 
	    far above the transition temperature ($T_{\rm COO}=231.5(5)$ K).
	    \label{width}}
\end{figure} 

Measurements in the high-temperature phase indicate 
that the structural correlations are very anisotropic. 
Figure \ref{width} reveals that the
in-plane correlations are nearly isotropic, with a ratio
$\xi_H/\xi_K\approx 0.63$, while the out-of-plane correlations
are much shorter. Above 255 K, the short-range order can
be considered two-dimensional, since the out-of-plane correlation length is on the
order of the interplane distance, $6.2$ {\AA} ($\approx c_t/2$).
The correlation lengths along the $H_o$ and $K_o$ directions cease to
decrease above that temperature and are
about $20$ {\AA} and $13$ {\AA}, respectively.

Even though $\xi_L$ remains finite in the ordered phase, the underlying
transition appears to be three-dimensional, and not two-dimensional, since the correlation lengths
for the three directions ($H_o$, $K_o$ and $L_o$) are approximately proportional to each other.
The situation in two-dimensional systems, for example the antiferromagnets K$_2$NiF$_4$ 
(Ref. \onlinecite{Birgeneau70}) or LaSrMnO$_4$ (Sec. III),
is very different since only two-dimensional fluctuations, that is, two-dimensional scattering rods, 
are observed. 
It is difficult to establish the correlation lengths close to the transition, in part
because of the chemical inhomogeneities discussed above, but also
because the estimate strongly depends on the shape of the resolution 
function. 
The development of full three-dimensional long-range order might be hindered due to
the presence of structural defects.
We note that it has been suggested that the three-dimensional nature of the charge/orbital order transition in 
La$_{0.5}$Sr$_{1.5}$MnO$_4$ results from an instability toward a structural distortion.
\cite{Onada02}

To complete this discussion of the charge/orbital order for $x=\frac{1}{2}$,
we present our findings for the effect of the x-ray probe on the charge-ordered phase.
At low temperatures, deep in the ordered phase, the scattering intensity at the superlattice 
position decreases but eventually levels off to a non-zero value when a relatively high x-ray flux is 
incident on the sample (about $10^{12}$ photons/second). This partial melting of the ordered
phase is demonstrated in the inset of Fig. \ref{opc}.
No such effect was observed when the 
incident flux was decreased by a factor of 300. A similar partial reduction of the 
superlattice intensity was 
found for most dopings in the charge/orbital ordered phase, and it resembles 
previous results 
observed for Pr$_{0.70}$Ca$_{0.30}$MnO$_3$ (Ref.
\onlinecite{Kiryukhin97}) and La$_{0.875}$Sr$_{0.125}$MnO$_3$ (Ref. \onlinecite{Kiryukhin99}).
This effect is likely related to the ``melting'' of the charge
order observed for both Pr$_{0.70}$Ca$_{0.30}$MnO$_3$ \cite{Miyano97} and
La$_{1.5}$Sr$_{0.5}$MnO$_4$ \cite{Ogasawara01} when samples were exposed to 
high-intensity visible light. These observations demonstrate that the 
low-temperature charge-ordered phase of La$_{1-x}$Sr$_{1+x}$MnO$_4$ is 
unstable when the material is exposed to intense electromagnetic radiation.

\begin{figure}[t]
  \centerline{\includegraphics[width=8.5cm]{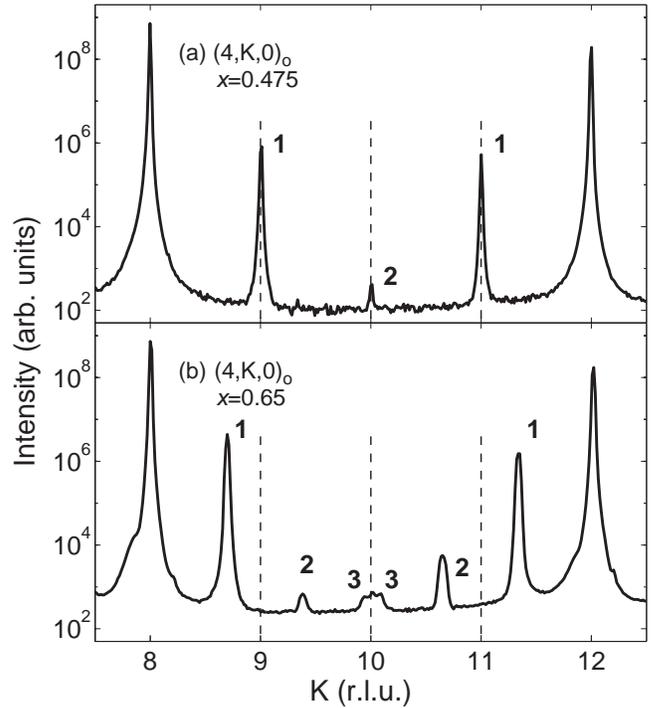}}
  \caption[Diffraction scans in the low-temperature phase for two dopings $x$]
      {X-ray diffraction scans of La$_{1-x}$Sr$_{1+x}$MnO$_4$ in the low-temperature 
	phase ($T=7$ K) along $(4,K,0)_o$ for (a) $x=0.475$ and (b) $x=0.65$. The vertical 
	dashed lines indicate the commensurate positions. 1, 2 and 3 label, respectively, the
	first, second and third harmonics of the low-temperature distortion. Note the 
	logarithmic intensity scale.}
      \label{incom4765}
\end{figure}

\subsection{Structural Properties for $0.45 \leq x < 0.70$}

\begin{figure}[t]
  \centerline{\includegraphics[width=9cm]{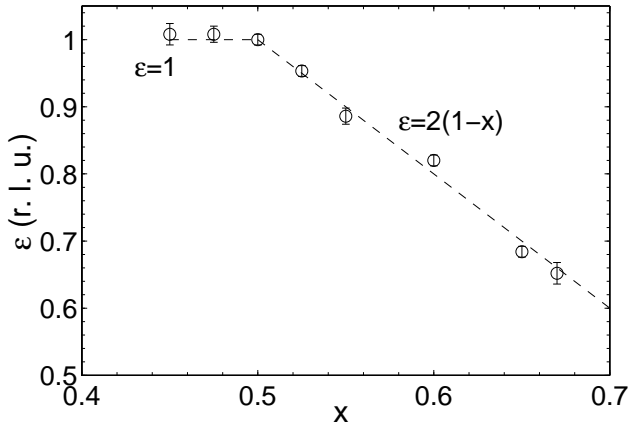}}
  \caption[Modulation wavevector as function of doping]
	  { Parameter $\epsilon$ of the low-temperature modulation wavevector 
	    $(0,\epsilon,0)_o$ as function of doping $x$ for La$_{1-x}$Sr$_{1+x}$MnO$_{4}$, determined
	    by synchrotron x-ray diffraction.
	    For $x \ge 0.5$, the value of $\epsilon$ is directly related to the $e_g$ electron density
	    $n_e = (1-x)$: $\epsilon = 2 n_e$.
		}
	  \label{incom}
\end{figure}

We observed superlattice peaks in all samples with $x>0.45$, up to $x=0.67$, the most Sr-rich
sample we were able to grow as a single crystal.
Figure \ref{incom4765} compares results for $x=0.475$ and $x=0.65$.
For $x > 0.50$, the superlattice modulation 
vector $\epsilon$ changes linearly with the $e_g$ electron density $n_e=1-x$, as shown in 
Fig. \ref{incom}. At $x=0.50$, the 
superlattice  modulation doubles the high-temperature structure (along the tetragonal base 
diagonal), and at $x=0.67$, it triples it. This linear dependence of the wavevector is 
similar to that observed in La$_{1-x}$Ca$_{x}$MnO$_3$ for $x>0.5$ \cite{Chen97} and, in 
particular, at $x=2/3$. \cite{Mori98b,Radaelli99,Fernandez99,Wang00} While at 
$x=1/2$ and $x = 2/3$ commensurate wavevectors are observed, the order is 
best understood, at all doping levels $1/2 \leq x \leq 2/3$, as a 
modulation whose period is a linear function of the doping $x$. 
We note that a section of one of our crystals with nominal doping of $x=0.65$ 
had an effectice doping level of $x \approx 0.69$, as judged from the measured
value of $\epsilon \approx 0.62$ (not shown in Fig. \ref{incom}).

\begin{figure}[t]
  \centerline{\includegraphics[width=8.8cm]{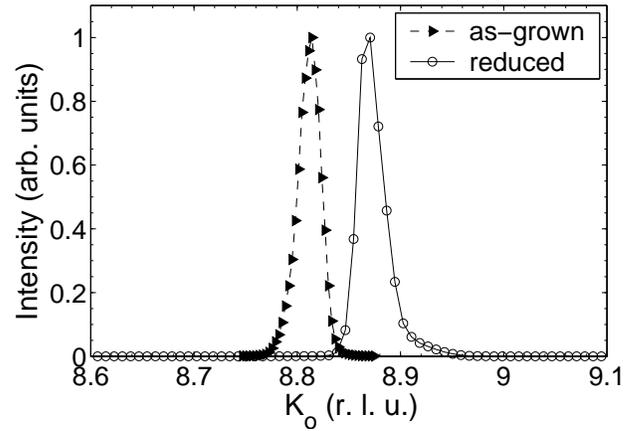}}
  \caption[Superlattice reflection for two $x=0.60$ samples with different oxygen 
    content]
	  { $K_o$-scans through the $(4,8+\epsilon,0)_o$ reflection for two samples with the 
	    same La/Sr content ($x=0.60$), but different oxygen stoichiometry. The first 
	    sample is as-grown while the second sample was annealed in argon 
	    ($P_{O_2}=10^{-5}$ bar) at 950$^\circ$C for 24 h.
	    \label{reduc}}
\end{figure}

The incommensurability of the modulation is also affected by the oxygen content. To 
establish this result, we compared the scattering from two pieces of the same crystal boule 
with a La/Sr ratio of 0.40/1.60. One piece was as grown, essentially quenched from a high
temperature in an environment with a relatively high oxygen partial pressure (5 bar).
The other sample was annealed for 24 h in a flow of argon with an oxygen partial pressure of 
$10^{-5}$ bar at a temperature of 950$^\circ$C. The second sample had a mass of 148.4 mg and 
exhibited a mass difference of 0.3(2) mg after the anneal, corresponding to a 
change in the oxygen content of 0.04(3). Figure \ref{reduc} demonstrates that the 
incommensurability (the deviation of the peak position from $K_o = 9$) 
decreased after the anneal. Using the linear relationship between the 
incommensurability and the nominal Mn valence established in 
Fig. \ref{incom}, one finds that the nominal 
Mn valence of the as-grown sample is 3.595 while that of the annealed sample is 3.56. In this 
simple ionic model, this result implies an oxygen content change of approximately 0.02, which is 
within the uncertainty of the mass measurement. Thus, the period of the modulation can be
linked to the $e_g$ electron density $n_e$. Because the superlattice modulation is directly
correlated with $n_e$, it is likely that the structural phase transition is driven by the 
ordering of the $e_g$ electrons.

In Fig.  \ref{incom4765},
second and third diffraction harmonics, much weaker than the primary,
are visible at $(0,\pm 2\epsilon,0)_o$ and $(0,\pm 3\epsilon,0)_o$. 
The relative weakness of the higher harmonics suggests that the structural distortion is 
essentially sinusoidal:
a pure sinusoidal modulation is described by a single Fourier component and exhibits weak higher 
harmonics.\cite{Axe80} In contrast, a nonsinusoidal modulation, especially one with sharp
discontinuities, would exhibit strong higher harmonics. For example, a square-wave 
modulation would exhibit strong odd harmonics; the intensity of the third harmonic would be
more than 10\% of that of the fundamental. Finally, we note that the widths of the superlattice 
peaks along $[1,0,0]_o$ and $[0,1,0]_o$ are comparable to those of the high-symmetry peaks, which 
implies that the low-temperature phase exhibits long-range order parallel to the Mn-O plane.

The nearly sinusoidal structural distortion, together with the linear variation of the 
modulation wavevector with doping, precludes any model in which the $e_g$ electron order is
too closely linked to the underlying cationic lattice, such as the bi-stripe
model,\cite{Mori98b} the topological scenario for stripe formation,
\cite{Hotta00} or the
the discommensurate-stripe model proposed for the single-layer 
nickelates. \cite{Yoshizawa00} A better description is given by a nearly sinusoidal 
structural distortion, probably  associated with a charge-density wave. The variation of 
the charge density, with equivalent Mn sites located as far apart as possible, is similar
to the ``Wigner-crystal'' arrangement of the $e_g$ electrons proposed for 
La$_{0.333}$Ca$_{0.667}$MnO$_3$ \cite{Radaelli99} (the term ``Wigner-crystal'' 
is somewhat unfortunate since the situation it 
describes is rather different from the classical Wigner crystal in which the electronic wave 
functions form a crystal independent of the nuclear lattice, as is observed in some 
semiconductors).

For $0.45\le x<0.50$, the modulation vector remains the same as for $x=0.50$. However,
the intensity of the superlattice peak 
increases as $x$ increases toward $x=0.50$. \cite{Larochelle01} The simplest explanation for this behavior is 
that the material separates into charge-ordered regions of 0.5 $e_g$ electrons per Mn site 
and disordered regions of approximately 0.55 $e_g$ electrons per Mn site. For compositions with
$x<0.45$, La$_{1-x}$Sr$_{1+x}$MnO$_4$ only exhibits short-range order. This will be 
discussed in Sec. V.

\begin{figure}[t]
  \centerline{\includegraphics[width=8.6cm]{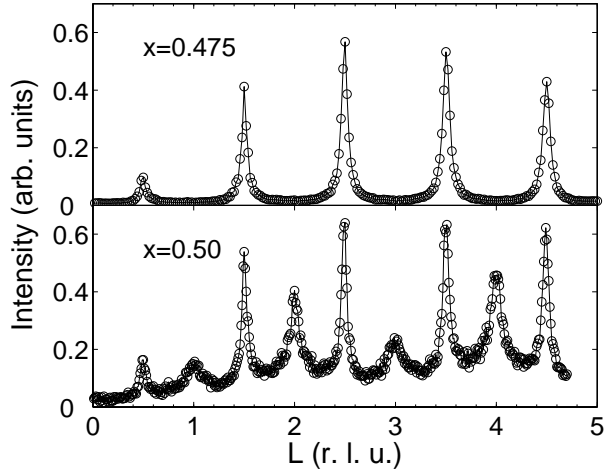}}
  \caption[ Neutron diffraction scans along $(1,0,L)_o$ for two different 
    samples: $x=0.475$ and $x=0.50$]
	  { Neutron diffraction scans along $(1,0,L)_o$ for 
	    $x=0.475$ and $x=0.50$. The measurements were taken at 
	    13 K and 7 K, respectively,
	    on the spectrometer BT7 with 13.4 meV
	    neutrons and collimations of 35$^\prime$-40$^\prime$-sample-25.8$^\prime$-open.
	    \label{magL}}
\end{figure}

\subsection{Magnetic Properties for $0.45 \leq x < 0.70$}

Using neutron scattering, we studied the magnetic order of several samples 
in the composition range $x\ge 0.45$. The doping levels of the samples were 
0.475, 0.50, 0.60 and 0.65. The first two samples showed N\'eel order below 
115 K, while only short-range antiferromagnetic order was observed in the two 
higher-doped samples. In all cases the magnetic-order wavevector was found to 
be commensurate with the lattice: $(1,0,L)_o$.

\begin{figure}[t]
  \centerline{\includegraphics[width=8.8cm]{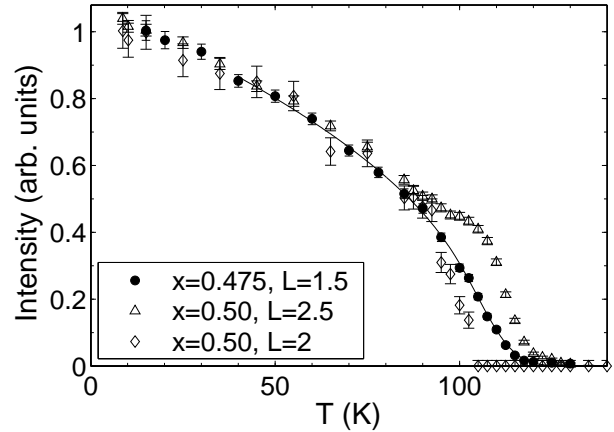}}
  \caption[Magnetic order parameter curves for samples with doping $x=0.475$ and $x=0.50$]
	  { Magnetic order parameter curves of the antiferromagnetic phase for samples with 
	    doping $x=0.475$ and $x=0.50$, as described in the text.
	    \label{neelCE}}
\end{figure}

For the two long-range-ordered samples, the observed magnetic structure is in good 
agreement with previous results\cite{Sternlieb96} for 
$x=0.50$, and it agrees with the CE structure. The authors of Ref. \onlinecite{Sternlieb96}
reported peaks corresponding to two stacking patterns along $[0,0,1]$.
The majority (minority) stacking pattern has magnetic peaks at half-integer (integer)
$L$ positions, corresponding to 
antiferromagnetic (ferromagnetic) next-NN planes, i.e., 
planes separated by a distance 
$c\approx 12.4$ {\AA}.
In the $x=0.475$ sample, we only observed peaks associated with 
the majority stacking pattern while, for $x=0.50$, we observed both patterns with a 
ratio slightly different from that in Ref. \onlinecite{Sternlieb96} (see Fig. \ref{magL}). The 
$x=0.50$ sample also exhibits significant diffuse scattering which gives rise to the higher 
background level seen in the bottom panel of  Fig. \ref{magL}.
Figure \ref{neelCE} demonstrates that  
the two sets of peaks in this sample have different temperature dependences,
with transitions of 105 K ($L$ integer) and 120 K ($L$ non-integer). 
In both samples, only short-range order was 
observed along $[0,0,1]$, while the peaks were resolution limited in the $H$-$K$ plane. The 
correlation lengths are $\xi_c=60(4)$ {\AA} for the majority stacking pattern (in both 
samples) and $\xi_c=30(4)$ {\AA} for the minority stacking pattern ($x=0.50$ only). The 
lengths do not vary below the N\'eel transition.

A power-law with Gaussian broadening (solid
line in Fig. \ref{neelCE}) gave a good fit of our $x=0.475$ data above 50 K, with
$T_N = 110(1)$ K, $\beta = 0.24(3)$, and 
a distribution of N\'eel temperatures with 
width (FWHM) of $15(1)$ K. The rounding is remarkably large, 
nearly an order of magnitude larger
than that typically found for the structural transition, and significantly 
larger than for the magnetic transitions in the low-doping regime.
One possible explanation for this observation is that, unlike for $x=0$,
the leading magnetic anisotropy is planar rather than uniaxial. It has been argued that
realistic, finite-size two-dimensional XY systems (or systems with a leading XY anisotropy)
should be characterized by a non-zero magnetization with effective exponent 
$\beta \approx 0.23$ and a significant effective rounding of the transition.
\cite{Bramwell93} Therefore, the value $\beta = 0.24(3)$ is consistent with the 
existence of a significant planar anisotropy.

\begin{figure}
  \centerline{\includegraphics[width=8.8cm]{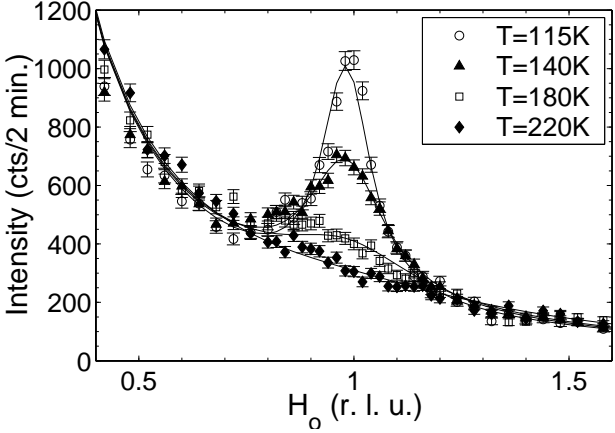}}
  \caption[Instantaneous magnetic structure factor measurement in the disordered phase for $x=0.475$]  
	  { Energy-integrating scans of the magnetic fluctuations in the disordered phase for
	    $x=0.475$ (two-axis neutron scattering mode) in the paramagnetic phase.
	    The lines are fits to a two-dimensional isotropic Lorentzian cross section 
	    convoluted with the instrument resolution.
	    The data were taken on the spectrometer BT7 with 13.4 meV initial energy
	    neutrons and collimations of 35$^\prime$-40$^\prime$-sample-25.8$^\prime$-open.
	    \label{MagSRO47}}
\end{figure}

\begin{figure}[t]
  \centerline{\includegraphics[width=8.8cm]{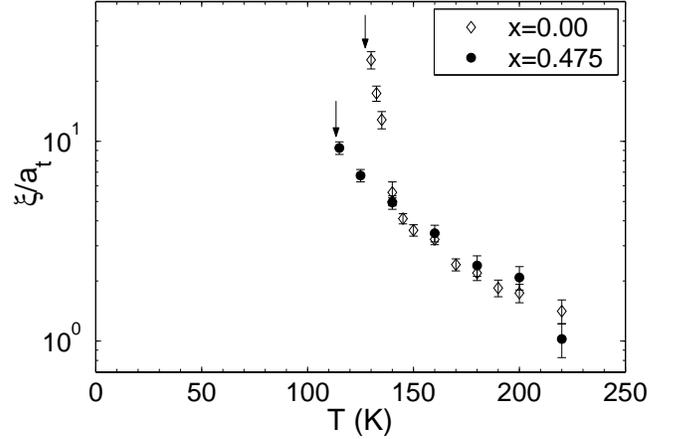}}
  \caption[Antiferromagnetic correlation length as function of temperature for $x=0.00$ and $x=0.475$]
	  {Instantaneous antiferromagnetic correlation length as function 
	    of temperature for $x=0$ and $x=0.475$. The downward arrows indicate 
	    the onset of long-range magnetic order.
	    \label{xi_0047}}
\end{figure}

Above the N\'eel transition, there exist significant two-dimensional antiferromagnetic
correlations for the $x=0.475$ sample
which are observable in the form of scattering rods. 
The scans shown in Fig. \ref{MagSRO47} were taken using a two-axis 
neutron spectrometer configuration in order to integrate over the energy of the 
fluctuations and to measure the instantaneous spin-spin correlations. 
The data are more limited than for $x \leq 0.15$ (Sec. III), and we were not able to
distinguish between in-plane (along [1,0,0]$_o$) 
and out-of-plane (along [0,0,1]$_o$) contributions. The correlation lengths
were extracted from the scans by assuming a single isotropic two-dimensional 
Lorentzian cross section convoluted with the instrument resolution. We note that,
due to the geometry of the experiment, the measuerment was not very sensitive to the fluctuations 
along [0,1,0]$_o$.
Figure \ref{xi_0047} compares the correlation lengths
for two compositions: $x=0.475$ and $x=0.00$. 
The latter result was discussed in Sec.
\ref{sect_gmag}. Interestingly, even though these two samples belong to two very
different magnetic phases, the high-temperature correlation lengths are indistinguishable. 
The two compounds thus probably have very similar values of the spin stiffness.
We observe no sign of a diverging length for $x=0.475$ even rather close to 
the onset of long-range magnetic order.

The discrepancy at low temperature can be explained 
by assuming that for $x=0.475$ there exist anisotropic $a$-$b$-plane spin correlations 
with $\xi_b >> \xi_a$ and 
a (possibly small) in-plane Ising anisotropy (in addition to an easy-plane anisotropy),
in contrast to 
the out-of-plane Ising anisotropy for $x=0.00$. 
While our data for $x=0.475$ establish that the magnetic scattering in the paramagnetic phase
is two-dimensional, they are not sufficient to conclude whether 
the nonequivalence of the $H_o$ and $K_o$ 
axes in the CE magnetic structure causes an anisotropy of
the in-plane magnetic correlation lengths; the two-axis scans 
primarily reflect the correlation in 
the $H_o$ direction, perpendicular to the ferromagnetic chains of the CE structure, 
and a two-dimensional isotropic 
Lorentzian cross section was assumed in the data analysis. 

The presence of short-range 
magnetic correlations between the N\'eel temperature and the charge/orbital order transition 
temperature has been argued to be necessary for the stabilization the charge/orbital ordered state. 
For example, it has been suggested that the correlations must be long along the 
ferromagnetic 
chains (along [0,1,0]$_o$) and short in the other directions. \cite{Solovyev03}
More experimental work is required to test these predictions.

\begin{figure}
  \centerline{\includegraphics[width=8.5cm]{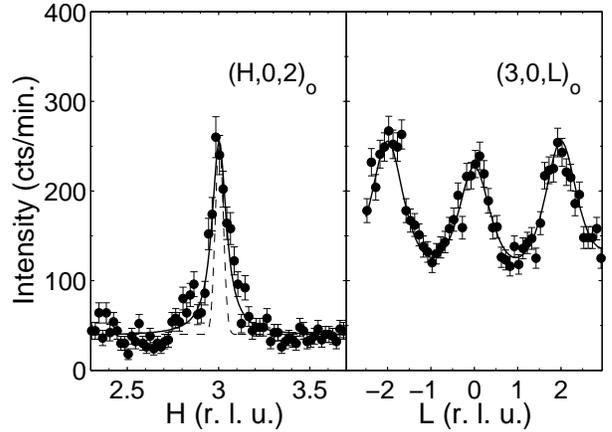}}
  \caption[Scans of the short-range magnetic correlations in a $x=0.60$ sample] 
	  { Triple-axis scans of the short-range magnetic correlations in a $x=0.60$ sample at low 
	    temperature ($T=9$ K). The dashed line in the left panel represents the 
	    instrument resolution. The magnetic correlations are short-range and essentially 
	    two-dimensional. The correlation length, as estimated from these triple-axis scans, 
	    is about 3.5 {\AA} along $[0,0,1]_o$ and 29(1) {\AA} along  $[1,0,0]_o$.
	    The data were taken on the spectrometer BT7 with 13.4 meV
	    neutrons and collimations of 35$^\prime$-40$^\prime$-sample-25.8$^\prime$-open.
	    \label{MagSRO60}}
\end{figure}

\begin{figure}
  \centerline{\includegraphics[width=8.5cm]{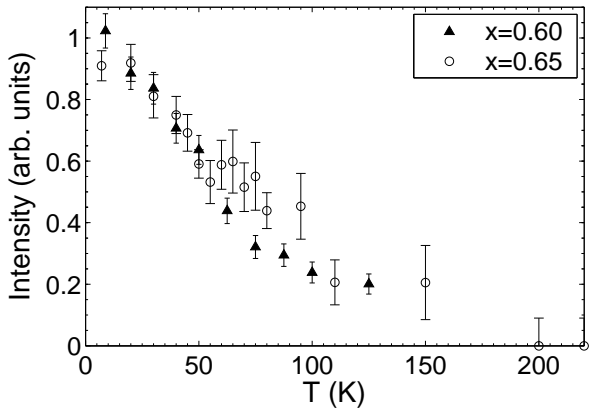}}
  \caption[Temperature dependence of the magnetic intensity in samples with doping
    $x=0.60$ and 0.65]
	  { Temperature dependence of the antiferromagnetic short-range peak intensity in 
	    samples with doping $x=0.60$ and 0.65.
    \label{tmag6065}}
\end{figure}

The two higher-doped samples, $x=0.60$ and $x=0.65$, only showed short-range magnetic order, 
essentially uncorrelated along [0,0,1] (the correlation length is less 
than the interplane distance). Along [1,0,0]$_o$, the peaks occur at the same 
commensurate positions as for $x=0.475$ and $x=0.50$. For both samples, 
the low-temperature correlation length along $[1,0,0]_o$, determined from 
triple-axis scans (see Fig. \ref{MagSRO60}), is only 29(1) {\AA}.
As demonstrated in 
Fig. \ref{tmag6065}, the peak intensity 
decreases approximately linearly with increasing temperature as the correlations become shorter. 
We observed a magnetic signal up to about 200 K. 

We note that the large x=0.60 and x=0.65 samples used in the neutron scattering measurement
were of somewhat lower quality than the small samples used for x-ray scattering
and those grown at lower strontium concentrations. The sample mosaics were on
the order of one degree. Also, the structural superlattice peaks were
broadened along the incommensurability direction, indicating the
existence of chemical inhomogeneities of about $\Delta x = 0.03-0.04$ (FWHM).
However, it is unlikely that a significant second
phase with doping x=0.50 exists in these x=0.60 and x=0.65 samples, since
no evidence of structural scattering
was observed at the commensurate position that could be associated with the
commensurate magnetic response.

The magnetic properties in this region of the 
phase diagram are quite different from those of La$_{0.33}$Ca$_{0.67}$MnO$_3$,
which has a very similar lattice distortion. The latter exhibits magnetic long-range order,
with a wavevector compatible with the tripling of the structural unit cell. \cite{Radaelli99}
It is thus quite surprising that the magnetic response of La$_{1-x}$Sr$_{1+x}$MnO$_4$
is commensurate with the underlying Mn-O lattice and, consequently, incommensurate with the 
structural distortion. It is probable that any long-range magnetic order would have to be 
commensurate with the lattice distortion modulation (i.e., incommensurate with the underlying
Mn-O lattice),
since it is unlikely that magnetic order and charge order are decoupled.
We speculate that for $x>1/2$
long-range magnetic order is forbidden in La$_{1-x}$Sr$_{1+x}$MnO$_4$ 
because of magnetic frustration effects, possibly associated with chemical disorder,
and due to the two-dimensional nature of the fluctuations.

\section{The Intermediate Region ($0.15<\lowercase{x}<0.45$)}
\label{sect_imr}

The long-range superstructure order observed at higher doping  disappears
rapidly as the doping level $x$ is lowered below $x=0.45$. In 
the intermediate doping regime, we find diffuse 
commensurate
scattering that is very similar to that in the 
disordered high-temperature phase for $x=0.50$. The peak position is slightly displaced: it is 
$(0.26,0.26,0)_t$ instead of $(\frac{1}{4},\frac{1}{4},0)_t$ for $x=0.50$. In the $x=0.50$ 
sample, both the long-range distortion of the low-temperature phase and the short-range 
distortion of the room-temperature phase are characterized by the commensurate wavevector
$(\frac{1}{4},\frac{1}{4},0)_t$. 

\begin{figure}
  \centerline{\includegraphics[width=8.5cm]{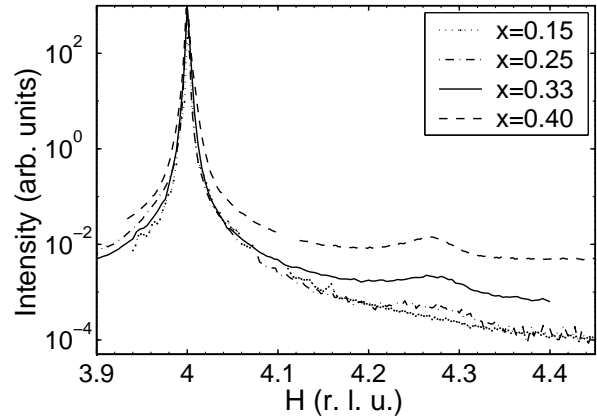}}
  \caption[Scans of a diffuse structural peak for $0.15\le x\le 0.40$] 
	  {Scans along $(H,8-H,0)_t$ for $x=0.40$, $x=0.33$ and $x=0.25$
	    revealing diffuse peaks
	    indicative of short-range order.
	    At lower doping
	    ($x=0.15$), however, no peak is observable. Small secondary-phase
	    peaks were found at $H=4.07$ ($x=0.15$ and $x=0.25$) and at
	    $H=4.12$ ($x=0.40$), but are not included in the figure.
	    \label{xless45}}
\end{figure}

The intensity of the diffuse structural scattering decreases rather rapidly at low doping, as can be 
seen in Fig. \ref{xless45}. The peak intensity decreases by an order of magnitude between 
$x=0.40$ and $x=0.33$ (this comparison should be considered qualitative
since absorption and extinction effects are considerable). The 
diffuse peak is still visible for $x=0.25$, but not for lower values of doping.

\begin{figure}
  \centerline{\includegraphics[width=8cm]{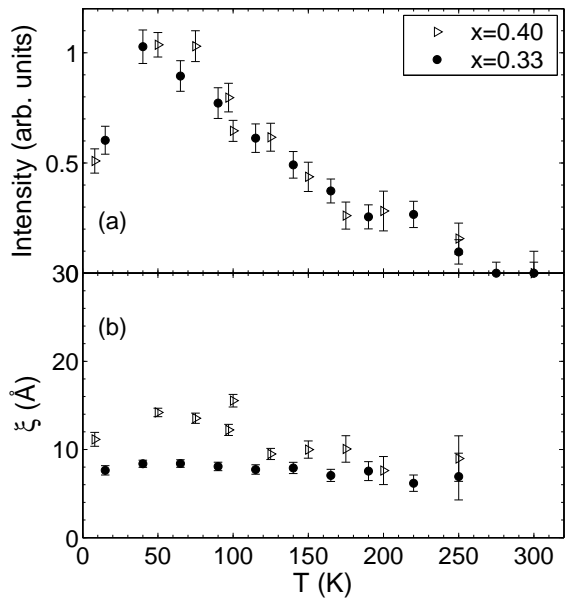}}
  \caption[Integrated intensity and structural correlation length for $x=0.33$ and 0.40.]
	  { (a) Integrated intensity and (b) correlation length along [1,0,0]$_o$ of the structural 
	    short-range order peaks
	    as function of temperature for two samples with intermediate 
	    doping: $x=0.33$ and $x=0.40$.
	    \label{quart3340}}
\end{figure}

As can be seen in Fig. \ref{quart3340}, the temperature dependence of both the intensity and 
the correlation length is weak. At low temperature, the integrated peak intensity decreases. 
Since these data were taken with a high x-ray flux, this decrease may be due to partial 
x-ray induced melting, similar to the effect observed in the long-range charge-ordered compounds 
(see Fig. \ref{opc}). 
The size of the correlated regions is comparable to those for $x=0.50$ in the 
disordered high-temperature 
phase (see Fig. \ref{width}). The onset temperature of the observed signal 
is about 250 K. 

\begin{figure}
  \centerline{\includegraphics[width=8cm]{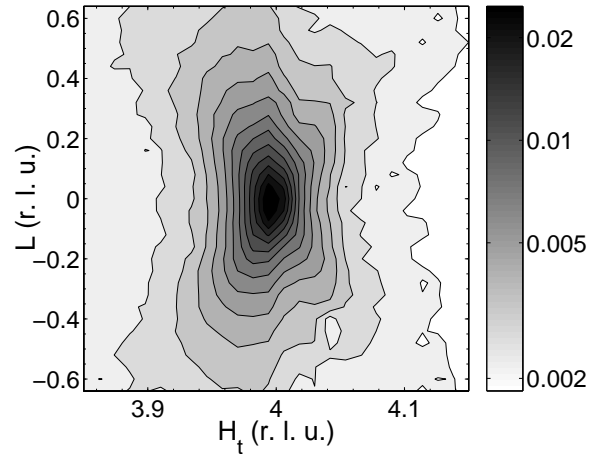}}
  \caption[Contour plot of the scattering intensity around the $(4,4,0)_t$ peak for a 
    $x=0.33$ sample]
	  { Logarithmic contour plot of the scattering intensity around the $(4,4,0)_t$ 
	    peak in a $x=0.33$ sample. The diffuse scattering is composed of thermal diffuse 
	    scattering (TDS) and Huang-like scattering due to Jahn-Teller distortions.
	    \label{contour}}
\end{figure}

\begin{figure}
  \centerline{\includegraphics[width=8cm]{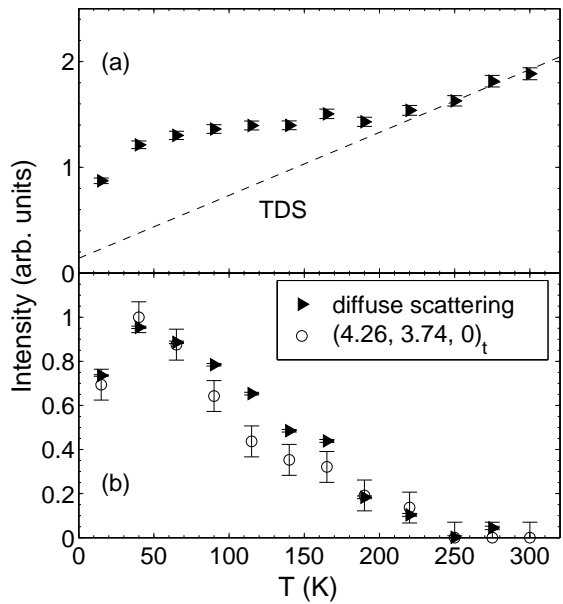}}
  \caption[Diffuse scattering intensity as function of temperature in a $x=0.33$ sample]
      { (a) Linear plot of the diffuse scattering intensity around the $(4,4,0)_t$ 
	peak as function of the temperature (for $x=0.33$). The straight line is an 
	estimate of the thermal diffuse scattering. (b) The Jahn-Teller component of the 
	diffuse scattering near $(4,4,0)_t$ 
 	scales linearly with the intensity of the broad peak 
	at position $(4.26,3.74,0)_t$ shown in Fig. \ref{quart3340}a.
	\label{diffquart}}
\end{figure}

Besides the diffuse scattering at the superlattice positions, significant
diffuse intensity was also observed in the ``tails" of the Bragg peaks, 
as can be seen from Fig. \ref{contour}.
We believe that the majority of this diffuse intensity results from Huang scattering
due to point defects, which can be calculated by modeling the local distortion of the
lattice due to these defects. \cite{Krivoglaz96}
Thermal vibrations distort a crystalline lattice, and these distortions of the perfect
lattice lead to a transfer of some of the scattering intensity from the Bragg peaks to their
tails.  The diffuse intensity in the tails due to thermal vibrations is commonly referred to as 
thermal diffuse scattering (TDS). A detailed analysis of the temperature dependence of the
diffuse intensity in the tails of the Bragg peaks of compounds with $0.25<x<0.45$
reveals an additional contribution.
At temperatures much higher than those corresponding to the energy of
the acoustic phonons in the momentum range of the measurement (essentially the whole
temperature range of the measurements reported here), a simplification of the Bose
factor that controls the phonon population yields a linear temperature dependence of
TDS. As shown in Fig. \ref{diffquart}, 
the subtraction of a linear TDS contribution
indeed reveals additional diffuse
scattering with an unusual temperature dependence.   
It can be seen from the bottom panel of Fig. \ref{diffquart} that the non-TDS part of the 
diffuse scattering scales linearly with the intensity of the 
correlated peak, and it should thus be related to these short-range distortions. Similar 
behavior has been found in colossal magnetoresistive double-layer and perovskite 
manganites. \cite{Vasiliu99,Shimomura99,Adams00} In these materials, short-range-ordered 
peaks and diffuse scattering around the Bragg points were observed to develop in the
paramagnetic insulating phase and the resulting distortions were named ``polarons."
These should not be seen as classical polarons, and we will refer to them 
here as correlated 
``nanopatches."

However, while the resistivity in the CMR materials correlates well with the 
scattering strength due to these nanopatches in the paramagnetic insulating phase, the resistivity 
of the single-layer manganite at these intermediate doping levels is dominated by other 
factors. The derivative $d ln(\rho)/d T^{-1}$ is constant in the case
of La$_{1-x}$Sr$_{1+x}$MnO$_4$, \cite{Bao96} 
which indicates that the resistivity $\rho$ is thermally activated. The conductivity 
is thus polaron-induced (the term polaron is used here with its usual meaning
in condensed matter physics). In contrast, the 
scattering strength of the correlated nanopatches varies much more 
slowly with temperature, and the resistivity due to the nanopatches varies almost linearly with 
temperature. Therefore, the nanopatches should not dramatically affect the transport properties of 
a polaronic insulator, as their contribution to the overall resistivity is minor. In contrast, 
the situation is very different in a material in which these correlated nanopatches dominate the 
resistivity, such as the CMR compounds. In these materials, a significant change in the 
density of these nanopatches, either via changes in temperature or applied magnetic field, 
dramatically changes the electrical conductivity from metallic to insulating.

A comparison of the short-range distortions of the single-layer ($n=1$), double-layer ($n=2$)
\cite{Vasiliu99,Argyriou02} and perovskite ($n=\infty$) \cite{Shimomura99,Adams00} manganites 
reveals an interesting fact: while all of them 
share a common CE-type distortion characterized by the 
wavevector $(\epsilon,\epsilon,0)_t$, the double-layer manganite also exhibits a distortion 
with a wavevector of $(\epsilon, 0,1)_t$, parallel to the Mn-O bond direction.

We also investigated the magnetic properties in this intermediate region of the 
phase diagram. As described in Sec. \ref{sect_gmag}, compounds above the 
$x_c\approx 0.115$ 
antiferromagnetic phase boundary exhibit short-range antiferromagnetic correlations
(doping levels $x=0.125$ and $x=0.15$). In addition, neutron scattering for $x=0.25$ reveals
only very weak two-dimensional scattering rods at 
wavevector $(1,0,L)_m$ (i.e., $(\frac{1}{2},\frac{1}{2},L)_t$) due to residual short-range
correlations of the antiferromagnetic phase at low doping. No correlated antiferromagnetic 
scattering was observed for $x=0.40$. Thus, we were unable to detect any 
CE-type antiferromagnetic correlations in the intermediate doping region.

Given the presence of a ferromagnetic phase in this doping range in the perovskite 
manganites, we also searched for short-range two-dimensional ferromagnetic 
correlations. Ferromagnetic correlations are significantly harder to detect than
antiferromagnetic correlations, because ferromagnetic scattering occurs at the same wavevectors
as the structural scattering. Furthermore, there exists a significant amount of diffuse 
scattering originating from the Bragg reflections along [0,0,1] because of the existence of 
stacking faults in layered compounds. Near $(2,0,L)_m$, we observed a weak neutron scattering signal that 
decreased with increasing temperature in the $x=0.15$ and $x=0.40$ compounds. This signal most likely 
originates from the Huang-like structural diffuse scattering discussed above.
Thus, we were unable to establish the presence of any ferromagnetic correlations
in this part of the phase diagram of La$_{1-x}$Sr$_{1+x}$MnO$_4$.

\section{Phase Diagram and Discussion}
\label{sect_pd}

\begin{figure}
  \centerline{\includegraphics[width=9.1cm]{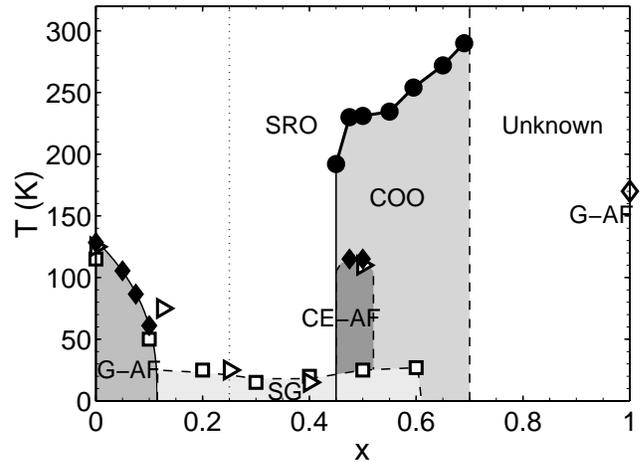}}
  \caption[Magnetic and structural phase diagram of La$_{1-x}$Sr$_{1+x}$MnO$_4$]
      { Magnetic and structural phase diagram of La$_{1-x}$Sr$_{1+x}$MnO$_4$.
	The data come from x-ray structural scattering (${\bullet}$), 
	neutron scattering ($\blacklozenge$ (this work) and 
	$\lozenge$ (Ref. \onlinecite{Bouloux81b})), 
	magnetometry ($\square$, Ref. \onlinecite{Moritomo95}) and muon spin rotation 
	($\triangleright$, Ref. \onlinecite{Baumann03}) measurements. 
	The abbreviations are G-AF: G-type 
	antiferromagnet, CE-AF: CE-type antiferromagnet, SG: spin glass, COO: 
	charge/orbital order phase, SRO: short-range charge and orbital order. The dotted
	line at $x=0.25$ does not denote a phase transition, but rather indicates the extent 
	to which SRO is visible. The dashed line at $x=0.70$ indicates the 
	approximate solubility limit
	of La$_{1-x}$Sr$_{1+x}$MnO$_4$. 
	\label{phasediag}
      }
\end{figure}

The x-ray and neutron scattering data presented in this paper, together with
previously published results, allow the construction of the magnetic and 
structural phase diagram for the single-layer manganite La$_{1-x}$Sr$_{1+x}$MnO$_4$, shown in
Fig. \ref{phasediag}. As seen in Sec. V, superstructural order exists above 
$x=0.45$, with a modulation wavevector that becomes incommensurate for $x>0.50$. The region 
$0.45 \le x < 0.50$ likely is a mixture of the 
ordered ($x=0.50$) and disordered ($x\approx 0.45$) structural phases. With the 
exception of the end compound Sr$_2$MnO$_4$ ($x=1$), the region with doping $x$ 
above 0.7 has not been studied since powders prepared by the solid state reaction 
method show two chemical phases \cite{Bao96} and single crystals could not be successfully grown. 
The lanthanum-rich part of the phase diagram is not presented
here, but the successful preparation of La$_{1.2}$Sr$_{0.8}$MnO$_4$ ($x=-0.2$) has been 
reported.\cite{Li00} 

Although the temperature dependence of the (anisotropic) superstructure correlations in the high-temperature
phase of La$_{0.50}$Sr$_{1.50}$MnO$_4$ indicate that the underlying transition is three-dimensional,
the correlation length perpendicular to the MnO$_2$ planes
remains finite ($\sim 200$ \AA) in the ordered phase, presumably due to the presence of defects.
The low-temperature structural symmetry ($B2mm$) is consistent with a Jahn-Teller-type
distortion due to orbital ordering. At room temperature for $x\ge 0.45$, and at all temperatures 
in the range $0.25\le x<0.45$, there exist only anisotropic short-range-distortions.
These distortions are quite similar to those observed in the paramagnetic insulating phase of 
the bilayer and perovskite CMR manganites. However, the resistivies of 
the single-layer manganite and of the CMR materials differ significantly. While the correlated 
distortions (``nanopatches") dominate the transport properties in the latter, the former is a 
conventional polaronic 
(or thermally-activated) insulator. 

At doping $x>0.50$, long-range superstructural order 
similar to that at $x=0.50$ is present, consistent with a 
nearly sinusoidal distortion.
However, this order is incommensurate with 
the lattice. The incommensurability is affected by both the lanthanum/strontium ratio 
and the oxygen content. For the as-grown samples studied here, we find the 
simple relationship $\epsilon=2(1-x)=2n_{e}$, linking the incommensurability to the $e_g$ 
electron density $n_e=(1-x)$.
These findings preclude models in which the $e_g$ electron order is
too closely linked to the underlying cationic lattice, such as the bi-stripe
model,\cite{Mori98b} the topological scenario for stripe formation,
\cite{Hotta00} or the
the discommensurate-stripe model proposed for the single-layer
nickelates.\cite{Yoshizawa00} 

Long-range magnetic order is present in only narrow ranges of doping, near $x=0$ and
$x=0.5$, and at $x=1$. In contrast to the structural modulation, the magnetic correlations 
remain commensurate 
with the Mn-O lattice at all doping levels. Short-range magnetic correlations, a remnant of 
the long-range commensurate magnetic order at $x<0.12$, are visible up to $x \approx 0.25$.
A spin-glass phase might exist in this intermediate range, as indicated
by magnetometry \cite{Moritomo95} and muon-spin-rotation \cite{Baumann03} measurements,
in agreement with the idea of a frustration-induced disappearance of the long-range 
antiferromagnetic order due to the random distribution of in-plane and out-of-plane
Jahn-Teller orbitals.

Comparing the results for La$_{1-x}$Sr$_{1+x}$MnO$_4$
with those of the perovskite and double-layer manganites,
\cite{Dagotto01,Salamon01} we conclude that the system with the most similar phase diagram is
La$_{1-x}$Ca$_{x}$MnO$_{3}$. \cite{Cheong99} The phase 
near $x=0.50$ is shared by both compounds, with 
similar patterns of Jahn-Teller distortions and in-plane magnetic structures. Also, in 
both cases the 
phase extends to high values of $x$, and it exhibits similar incommensurate 
superstructures.
\cite{Chen97} Both compounds are antiferromagnets near $x=0$, although 
(La,Ca)MnO$_3$ is an A-type antiferromagnet while the spin order of 
(La,Sr)$_2$MnO$_4$ is closer 
to that of the C- and G-type perovskite magnetic structure. The Jahn-Teller distortion around 
the Mn atoms is also different in the two systems, \cite{vanElp00} as one would expect,
considering the strong spin/orbital coupling in the manganites. \cite{Tokura00} In the 
intermediate doping region $0.17\le x < 0.50$, (La,Ca)MnO$_3$ and (La,Sr)$_2$MnO$_4$ are 
both paramagnetic insulators at high temperature with short-range-correlated 
structural distortions.
\cite{Adams00} However, while the former becomes a ferromagnetic metal at low temperature,
the latter remains a paramagnetic insulator with very short two-dimensional magnetic correlations. 
This difference results in the absence of CMR for the single-layer manganite. 
We believe that the lack of 
a ferromagnetic phase in La$_{1-x}$Sr$_{1+x}$MnO$_4$ is the result of its 
quasi-two-dimensional structure and the very weak magnetic coupling along [0,0,1].

The single-layer manganite La$_{1-x}$Sr$_{1+x}$MnO$_4$ 
has three long-range-ordered low-temperature phases at dopings
of $x=0$, $x=0.50$ and $x=1$. The two compounds with integer nominal manganese valences, 
i.e., Sr$_2$MnO$_4$ with Mn$^{4+}$ and LaSrMnO$_4$ with Mn $^{3+}$, exhibit long-range 
G-type antiferromagnetic order with an Ising anisotropy and magnetic moments 
along $[0,0,1]$. The $x=0.50$ compound shows CE-type spin order with the 
moments aligned within the $a-b$ plane; it is thus very different from the two end compounds.
The phase diagram can be understood by considering the magnetic interactions
between manganese sites with formal valences of 3+ and 4+. The introduction of Mn$^{4+}$
sites into the Mn$^{3+}$ matrix of LaSrMnO$_4$ first frustrates the antiferromagnetic 
order by randomly introducing NN ferromagnetic interactions between
Mn$^{3+}$ and Mn$^{4+}$ sites. Near $x=0.25$, small correlated nanopatches
with ordered
Mn$^{3+}$ and Mn$^{4+}$ sites appear. A phase transition to large clusters of 
this mixed-valence order only occurs for $x\sim 0.45$.
However, this state is never fully long-range ordered structurally or magnetically
perpendicular to the Mn-O planes. Above $x=0.50$, the persistance of structural order 
likely indicates that the Jahn-Teller distorted Mn$^{3+}$ sites form a relatively
well-ordered pattern, with a modulation wavevector that increases linearly with doping.
Finally, at $x=1$, 
antiferromagnetic interactions between nearest neighbor Mn$^{4+}$ sites result in a 
long-range antiferromagntic state. 

In summary, we have carried out a detailed x-ray and neutron scattering study of the 
single-layer manganese oxide La$_{1-x}$Sr$_{1+x}$MnO$_4$, a structural homologue
of the well-studied high-temperature superconductor 
La$_{2-x}$Sr$_x$CuO$_4$, and the $n=1$ end-member of the Ruddlesden-Popper series
La$_{n(1-x)}$Sr$_{nx+1}$Mn$_n$O$_{3n+1}$.
Our results for the doping dependence of the magnetic and structural properties give
a refined understanding of the phase diagram of this material. We hope that 
these results
will help
guide the development of 
sophisticated theories for
La$_{1-x}$Sr$_{1+x}$MnO$_4$.
Since the complex CE phase observed near $x=1/2$ is also relevant to the physics of the
colossal magnetoresistance manganites ($n=2$ and $n = \infty$), our
detailed results for this part of the phase diagram of La$_{1-x}$Sr$_{1+x}$MnO$_4$
might furthermore contribute to a deeper understanding of the CMR phenomenon.

We would like to thank J.E Lorenzo, J.P Hill, and N. Nagaosa for helpful comments.
SSRL is supported by the DOE
Office of Basic Energy Sciences, 
Divisions of Chemical Sciences and Materials Sciences.
The work at Stanford University was furthermore supported
by the DOE under Contracts No.
DE-FG03-99ER45773 and No. DE-AC03-76SF00515, and by NSF CAREER
Award No. DMR9985067.

\bibliography{prb}
\end{document}